\documentclass[ twocolumn]{aastex631}

\newcommand{\psrone}{J0036$-$1033}
\newcommand{\psrtwo}{J0026$-$1955}

\newcommand{\psrfour}{J1357$-$2530}

\newcommand{\psrfive}{J0452$-$3418}
\newcommand{\dmu}{\ensuremath{\rm pc\,cm^{-3}}}
\newcommand{\uJybm}{\ensuremath{\mu {\rm Jy\,beam}^{-1}}}
\newcommand{\rmu}{\ensuremath{\rm rad\,m^{-2}}} 
\newcommand{\code}[1]{\textsc{#1}}

\usepackage{hyperref}
\usepackage[flushleft]{threeparttable}
\usepackage{subfigure}
\usepackage{amsmath}
\usepackage{tabularx}
\usepackage{multirow}

\begin{document}

\title{Discovery and follow-up of a quasiperiodically nulling and sub-pulse drifting pulsar with the Murchison Widefield Array}

\author[0000-0001-5561-1325]{Garvit Grover}
\affiliation{International Centre for Radio Astronomy Research, Curtin University, Bentley, WA 6102, Australia}

\author[0000-0002-8383-5059]{N. D. Ramesh Bhat}
\affiliation{International Centre for Radio Astronomy Research, Curtin University, Bentley, WA 6102, Australia}

\author[0000-0001-6114-7469]{Samuel J. McSweeney}
\affiliation{International Centre for Radio Astronomy Research, Curtin University, Bentley, WA 6102, Australia}

\author[0000-0001-6840-4114]{Christopher P. Lee}
\affiliation{International Centre for Radio Astronomy Research, Curtin University, Bentley, WA 6102, Australia}

\author[0000-0001-8845-1225]{Bradley W. Meyers}
\affiliation{International Centre for Radio Astronomy Research, Curtin University, Bentley, WA 6102, Australia}

\author[0000-0001-7509-0117 ]{Chia Min Tan}
\affiliation{International Centre for Radio Astronomy Research, Curtin University, Bentley, WA 6102, Australia}

\author[0000-0002-6631-1077]{Sanjay S. Kudale}
\affiliation{National Centre For Radio Astrophysics - Tata Institute Of Fundamental Research, Pune 411 007, India}




\begin{abstract}
The phenomenon of pulsar nulling, where pulsars temporarily and stochastically cease their radio emission, is thought to be indicative of a `dying' pulsar, where radio emission ceases entirely. 
Here we report the discovery of a long-period pulsar, PSR~\psrfive, from the ongoing Southern-sky MWA Rapid Two-meter (SMART) pulsar survey. 
The pulsar has a rotation period of ${\sim}$1.67\,s and a dispersion measure of 19.8\,\dmu, and it exhibits both quasi-periodic nulling and sub-pulse drifting. 
Periodic nulling is uncommon, only reported in $<1$\% of the pulsar population, with even a smaller fraction showing periodic nulling and sub-pulse drifting. 
We describe the discovery and follow-up of the pulsar, including a positional determination using high-resolution imaging with the upgraded Giant Metrewave Radio Telescope (uGMRT), initial timing analysis using the combination of MWA and uGMRT data, and detailed characterisation of the nulling and drifting properties in the MWA's frequency band (140-170\,MHz).  
Our analysis suggests a nulling fraction of 34$\pm6$\% and a nulling periodicity of 42$^{+1.5}_{-1.3}$ pulses. 
We measure the phase ($P_2$) and time modulation ($P_3$) caused by the sub-pulse drifting, with an average $P_2$ of 7.1$^{+26.3}_{-3.1}$ degrees and a $P_3$ of 4.8$^{+1.5}_{-0.9}$ pulses. 
We compare and contrast the observed properties with those of other pulsars that exhibit sub-pulse drifting and quasi-periodic nulling phenomena, and find that the majority of these objects tend to be in the `death valley' in the period-period derivative ($P$-$\dot{P}$) diagram. 
We also discuss some broader implications for pulsar emission physics and the detectability of similar objects using next-generation pulsar surveys.

\end{abstract}




\section{Introduction} \label{sec:intro}

Pulsars are known for their periodic emission. However, some pulsars occasionally cease their emission for seemingly random durations; this is referred to as `nulling' \citep{Backer1970}. The nulling phenomenon has been observed in ${\sim} 8\%$ of the pulsar population \citep{Sheikh2021}, yet the origins remain unknown. 
Nulling has been suspected to be indicative of an ageing, `dying' pulsar (a pulsar ceasing emission permanently), in which case the nulling fraction (i.e. the fraction of time the pulsar spends in its nulling state) 
can be expected to correlate with period and characteristic age, as proposed by, e.g., \citet{Ritchings1976} and \citet{Biggs1992}.
 
However, some recent studies find no strong correlation with either characteristic age or period \citep{Sheikh2021, Rankin1986, Wang2007, Anumarlapudi2023}, and hence no strong link between pulsar `death' and nulling has yet been made \cite[for a comprehensive summary of recent work in this area, see][]{Sheikh2021}. 


Nulling behaviour between pulsars can be very diverse \citep[see for example Figure 2 of][]{Wang2007}. As we do not understand the core mechanism behind nulling, it is unclear whether the same process causes nulling in all pulsars. 
There have been efforts to understand the population of nulling pulsars, by considering the sub-classes with low and high nulling fractions; specifically, those with nulling fractions $>40\%$ and the other $<40\%$ \citep{Konar2019}. However, more recent studies suggest a lack of significant distinction between the two groups \citep{Sheikh2021}.  
It is therefore also useful to analyse pulsars individually to explore and understand their nulling behaviour.

Pulsars generally tend to null stochastically with no obvious pattern; however, about ${\sim}10\%$ of nulling pulsars show periodic or some quasi-periodic nulling \citep[e.g.,][]{Basu2020, Basu2017, Anumarlapudi2023, Herfindal2007, Herfindal2009}. 
This may be an underestimate considering the paucity of systematic studies that specifically investigate periodicity in nulling behaviour. 
Studies of quasi-periodic nulling are particularly valuable because the periodicity itself can then be used to constrain models of the nulling mechanism.
If the periodicity in nulling can be meaningfully characterised, it can then be compared to other physical properties of pulsars to gain vital clues.  
Although attempts to explain periodic nulling have not been conclusive, there have been some suggestions that quasi-periodicity in nulling is possibly linked to sub-pulse drifting \citep{Herfindal2007, Herfindal2009}. 

Sub-pulse drifting is a phenomenon whereby the peak of the pulse shifts in phase as the pulsar rotates, which results in the appearance of \textit{drift bands} (diagonal tracks) when the data are represented in the form of single-pulse stacks. The sub-pulse drifting behaviour of pulsars can be characterized using the quantities $P_2$ and $P_3$, and in some cases also by a qualitative categorization of finite pulse sequences into so-called `drift modes'. In pulse stacks, where the pulse number is plotted against the pulse phase, $P_2$ is the spacing between consecutive bands of sub-pulses in the pulse phase direction, and $P_3$ is the spacing between consecutive bands of sub-pulses in the pulse number direction. Additionally, different `drift modes' can be defined as sets of drifting patterns that are characterised by different values of $P_3$.

Sub-pulse drifting is often explained in terms of the carousel model \citep{Ruderman1975}, which invokes the presence of multiple emission regions, or `beamlets', rotating around the polar cap of the neutron star. Many of the nulling pulsars also tend to show sub-pulse drifting (see \citealt{Basu2020}, and references therein). \cite{Basu2020} compiled a list of all such known objects, and found 15 pulsars that show quasi-periodic nulling and sub-pulse drifting.
This may suggest a connection between the nulling and drifting phenomena.

Here we present the discovery of a 
long-period ($P\sim$ 1.67\,s) pulsar, PSR~\psrfive, which shows both quasiperiodic nulling and sub-pulse drifting. 
It displays sub-pulse drifting behaviour
with a consistent drift rate and a singular drift mode. This pulsar was discovered in the first pass processing of the Southern-sky MWA Rapid Two-meter (SMART) pulsar survey \citep{smartone,smarttwo}. The discovery of the pulsar, initial follow-up, and the observations used are described in Sections \ref{sec:obs}. The methods, and results from the nulling and sub-pulse drifting analysis, and other related analyses are described in Section \ref{sec:analysis}. 
In Section \ref{sec: disc}, we discuss the sub-pulse drifting and nulling analysis and prospects of future discoveries and our findings are summarised in Section \ref{sec:con}.

\begin{figure*}[ht]
    \centering
    \includegraphics[width=\linewidth]{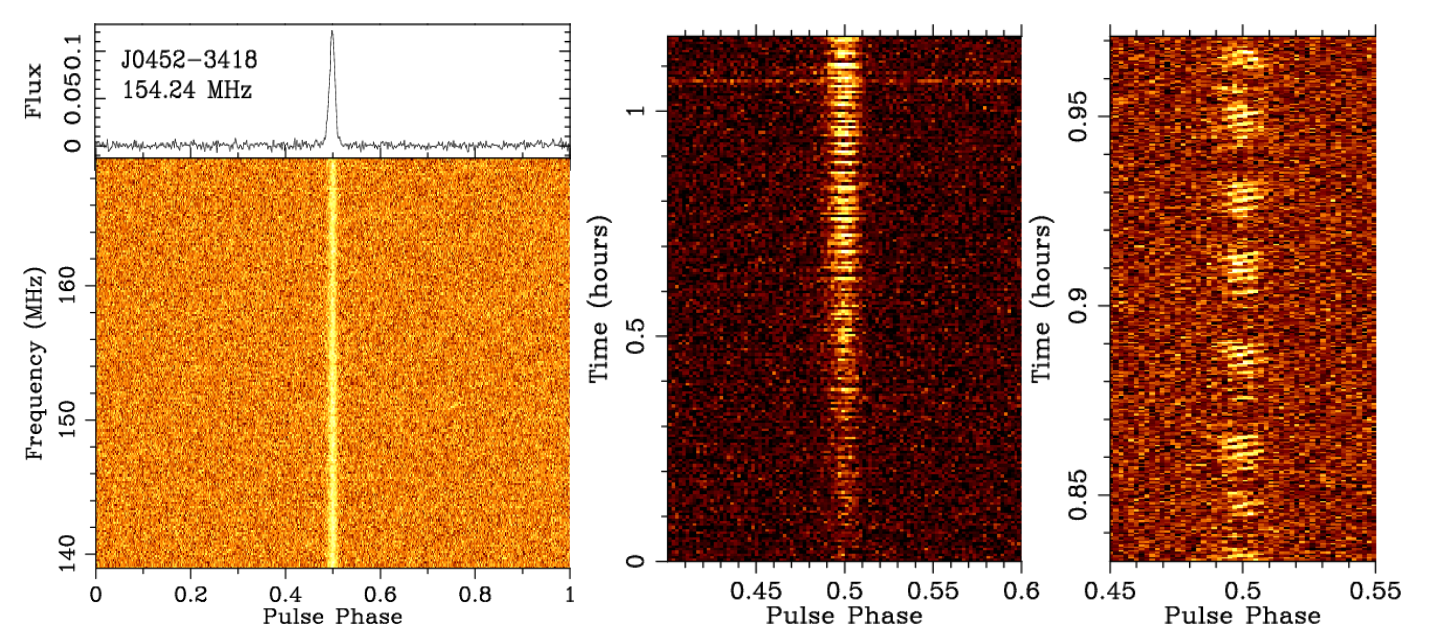}    
    \caption{The integrated profile across the MWA band (the left panel) and the time vs phase plot of PSR~\psrfive{} in observation R04. The middle panel shows a pulse stack over a $\sim70$\,min of data. The pulsar is drifting into the 
    telescope's primary beam, hence  appears to become brighter with time. The rightmost panel shows a zoomed-in version of the middle panel, and highlights sub-pulse drifting as well as a periodic nulling.}
    \label{fig:profile+stack}
\end{figure*}
 
\section{Discovery and Follow-up\label{sec:obs}}
\subsection{Survey Background}

The SMART survey is an all-sky pulsar search project that exploits low frequencies and the large field of view of the Murchison Widefield Array (MWA; \citealt{Tingay2013}) to survey the southern sky for pulsars and transient signals \citep{smartone,smarttwo}. 
It employs long dwell times (80 minutes per pointing) and records data in the form of raw voltages (rather than channelised detected data). 
The survey is conducted in the 140-170 MHz, 
and the survey design is such that there are overlapping observations over the whole southern sky ($\delta < 30^{\circ}$), making it an optimal data set for detecting pulsars with long null durations; e.g. the discovery of PSR~\psrtwo{} by \cite{Mcsweeney2022}. The survey uses the compact configuration \citep{2018PASA...35...33W} of the MWA which has a tied-array beam size of $\sim$24\,arcminutes. 
The large field of view ($\sim$600\,${\rm deg^2}$ at 150\,MHz) and the voltage recording means the entire southern sky can be surveyed in  ${\sim}100$ hours  \citep{smartone}.
However, the use of low frequencies also means a very large number of DM trials, and as a result, the processing generates millions of candidate signals, which must then be scrutinised after filtering through machine learning software. The filtered candidates are stored on a web app hosted by DataCentral\footnote{\url{https://apps.datacentral.org.au/smart/}}.  The candidates are then rated via visual inspection and only highly-rated candidates are followed up. PSR~\psrfive\ was discovered in the initial stages of this development. 

\subsection{Detection and initial follow-up}

PSR \psrfive{} was discovered through its multiple candidates in the SMART observation R11 (see Table \ref{tab:obssummary} for details). This 
is one of the observations from the 2019 observing campaign (shown in red in Figure 1 of \citealt{smartone}). The candidates were clustered in the `DM-period-sky' parameter space, given their closeby values of dispersion measure (DM),  period and position on the sky. The observations for the SMART survey are designed such that their observing regions overlap with neighbouring observations. Hence there can be more than one observation in which a given source may be detected. Candidates of similar DM, period and locations were later also found in two other observations, R09 and R04.

\begin{table*}[ht]
    \centering
    \caption{Summary of the observations used for \psrfive{}}
    \begin{threeparttable}
    
    \bgroup
    \def\arraystretch{1.5}%
    \begin{tabular*}{\textwidth}{c@{\extracolsep{\fill}}c c c c c c c c c   }
    \toprule
    
         \multirow{2}{*}{Observation} & RA/Dec & \multirow{2}{*}{Telescope} & Periods & \multirow{2}{*}{MJD} &\multirow{2}{*}{S/N} & Flux Density  & Center Freq  & Bandwidth \\
         &(deg)&&observed&&&(mJy)&(MHz)&(MHz)\\
         \hline
         R04& 59.6/-40.5\tnote{\textit{d}} &MWA & 2522&58757& $\sim 30$\tnote{\textit{a}} &$13\pm6.5$\tnote{\textit{g}}&155& 30.72\\
         R09 &76.2/-26.7\tnote{\textit{d}}&MWA & 301&58792& $\sim 9$\tnote{\textit{a}} &$7.2\pm3.6$\tnote{\textit{g}}&155& 30.72\\
         R11 & 92.6/-40.5\tnote{\textit{d}} & MWA & 361&58806& $\sim 11$\tnote{\textit{a}}&$9.2\pm4.6$\tnote{\textit{g}} &155& 30.72\\
         GM1& 73.0425/-34.3147 &GMRT & N/A &60065 & N/A & $0.32\pm0.16$ &400&200  \\
         GM2& multiple pointings\tnote{\textit{f}} &GMRT &${\sim}1100$ &60077 &$\sim20$\tnote{\textit{c}}&$3.57\pm0.35$&400&200\\
         GM3& 73.003/-34.3117& GMRT&${\sim}1100$ & 60084&$\sim60$\tnote{\textit{c}}&$4.56\pm0.61$&400&200\\
         GM4&73.003/-34.3117 & GMRT&${\sim}1100\times 4$\tnote{\textit{e}} &60174 &$\sim166$\tnote{\textit{ab}}&$0.67\pm0.05$&400&200\\
         GM5 & 73.003/-34.3117 & GMRT&${\sim}1100\times 4$\tnote{\textit{e}} &60178 &$\sim158$\tnote{\textit{ab}}&$0.95\pm0.13$&400&200\\
         \hline
         \end{tabular*}
        \egroup
        \begin{tablenotes}
        \small
        
        \item[\textit{a}] The S/N was measured using \code{psrchive} integrated profile measurements
        \item[\textit{b}] These S/N are influenced by the inclusion of RFI
        \item[\textit{c}] These S/N were obtained from \texttt{pdmp}
        \item[\textit{d}] The RA/Dec reported here are the center RA/Dec of the whole drift scan observation. The standard for reporting the RA/Dec of MWA observations changed $\sim$2019, where the standard now is to report the initial center of the beam rather than the center of the whole drift scan observations.
        \item[\textit{e}] The GMRT observations are taken in 30\,min chunks, the periods here represent the total period collected over the whole 2\,hrs
        \item[\textit{f}] Multiple pointings in an effort to localise the pulsar position (see text for details); pulsar detection was made toward Source D in Fig. 2 (RA/Dec: 72.9996/-34.3506).   
        \item[\textit{g}] These flux measurements were obtained using the radiometer equation with system temperature = 300\,K and gain $= 0.5$\,K/Jy \citep{smartone}.
        \end{tablenotes}
    \end{threeparttable}
    \label{tab:obssummary}
\end{table*}

The candidates were produced using only the first 10\,minutes of the full 80\,minute observations. The strongest detection significance of the candidates reported by \code{presto} was $25\sigma$ at a position of RA 04h51m22s Dec $-34^\circ 21^{\prime}11^{\prime\prime}$. For initial follow-up, the full 80 minutes of R04 (42\,TB) were processed. The source position remained within the MWA primary beam for a longer duration in observation R04 (${\sim}70$\,min) than in the other two observations (R09 and R11). Using the \texttt{pdmp} algorithm in the \code{psrchive} package, which uses a brute force grid search to extract basic parameters, we measured an initial period of ${\sim}1.67$\,s and a 
DM of ${\sim}19$\,\dmu. The detection from observation R04 is shown in Figure \ref{fig:profile+stack}; as evident from this figure, the pulse stacks clearly display the phenomena of both sub-pulse drifting and quasi-periodic nulling.


We used a strategy similar to that was used by \cite{psrone} and \cite{smarttwo} for obtaining an initial localisation of the pulsar using SMART data. This method exploits the fact that the MWA data are stored as raw voltages, thereby allowing us to beamform at multiple nearby positions (in the form of a dense grid) and compare the significance of the 
resultant detections with that expected from the predicted tied-array beam responses. Applying this method to observation R11, we determined a position of RA 04h52m10s, Dec $-34^\circ 18^{\prime}52^{\prime\prime}$, with a 1-$\sigma$ uncertainty of ${\sim}3^{\prime}$. 
Ideally, we would perform this procedure for every position to account for ionospheric shifts as they can cause large offsets in the apparent position of the source \citep{2022PASA...39...20S}. 
Re-beamforming data from observation R04 towards this updated position increased the \code{presto}-reported significance from $25\sigma$ to ${\sim}35\sigma$. This self-localisation was sufficiently precise to consider following up with other telescopes 
 to further improve the localisation.

\subsection{uGMRT imaging for localisation}\label{sec:localise}

As described in \citet{smarttwo}, follow-up with sensitive high-resolution interferometers like the upgraded Giant Metrewave Radio Telescope (uGMRT) is an integral part of the follow-up strategy for the SMART survey. The benefits of this, particularly precise localisation possible via high-resolution imaging, was convincingly demonstrated for early pulsar discoveries by \citet{psrone} and \citet{smarttwo}, where uGMRT imaging in the Band 3 (300-500 MHz) and Band 4 (550-750 MHz) enabled sub-arcsecond level localisation for PSRs~\psrone{} and \psrtwo. We adopted a similar strategy for PSR \psrfive, though in this case, observations were restricted to the Band 3 range, a choice that was largely driven by the relatively poor localisation achieved thus far with the MWA (owing to the detections being made in array's compact configuration), and partly motivated by further detailed characterisation of the pulsar's drifting and nulling characteristics. Multiple observations were made in 2023 May and 2023 August (GTAC Cycle 44; Observations GM1 - GM5; see Table 1) and data were recorded at a time resolution of 655.56\,$\mu$s and 2048 frequency channels across the 200\,MHz recording bandwidth (see Table 1 for further details). 

\begin{figure*}[t]
\centering
\includegraphics[width=0.480\linewidth]{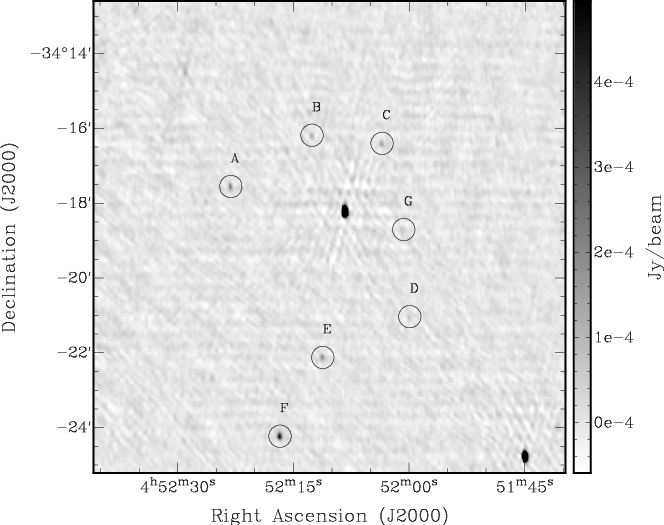}
\hspace{10pt}
\includegraphics[width=0.480\linewidth]{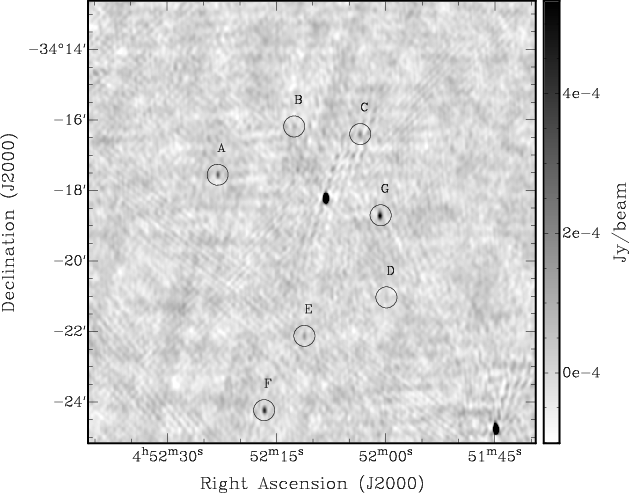}
\caption{uGMRT images of the PSR~\psrfive{} field in Band 3 (300-500\,MHz), obtained from observations made on 1 May 2023 ({\it left}) and 13 May 2023 ({\it right}); the integration times are 48\,min and 30\,min, respectively, and the corresponding RMS sensitivities are 
$\approx$166\,\uJybm{} and $\approx$345\,\uJybm, respectively. 
Imaging of the initial (poorly localised) pulsar position derived from MWA analysis revealed multiple compact sources (positions A-F) as shown in the figure, though a new source appeared in later imaging (right panel) turned out to be the pulsar (position G). There is a hint of this
source in the first imaging (left panel), though this initial detection is very marginal (2$\sigma$), with an estimated flux density of $0.32\pm0.16$\,mJy, whereas the detection significance is much higher (10$\sigma$) in the later observation, with an estimated flux density of $3.57\pm0.35$\,mJy. 
}
\label{fig:psrfiveimage}
\end{figure*}

The observing setup and analysis procedures employed are quite similar to those described in \cite{psrone} and \cite{Mcsweeney2022}, wherein concurrent imaging and phased-array (PA) observations were made around the initial pulsar position derived from the MWA follow-up. However, a more conservative strategy was adopted with the choice of PA beams; specifically, owing to the relatively more uncertain (${\sim} 3^{\prime}$) MWA position, one of the phased array (PA) beams was formed from only antennas within the central 1\,km $\times$ 1\,km region (thence a beam size ${\sim}3.8^{\prime}$ at 400 MHz). Following an initial non-detection of the pulsar, visibility data (recorded at a resolution of 2.67\,s) of observation GM1 were imaged, which revealed multiple compact sources within a ${\sim}5^{\prime}$ radius around the initial pulsar position, as shown in Fig.~\ref{fig:psrfiveimage} left panel (denoted as A to F), where the RMS sensitivity is 207\,\uJybm{} over the 48 min integration. This can be compared to the case of \psrfour{} \citep[see Section~4.1.4 of][]{smarttwo}. Subsequently, phased-array beam observations were made on each of these six sources (where the PA beam was formed from all working antennas out to $\sim$6-8\,km baselines). A detection with a modest significance (S/N$\sim$20) was made in the PA beam toward position D. 

\begin{figure}[ht!]
    \centering
    \includegraphics[width=\linewidth]{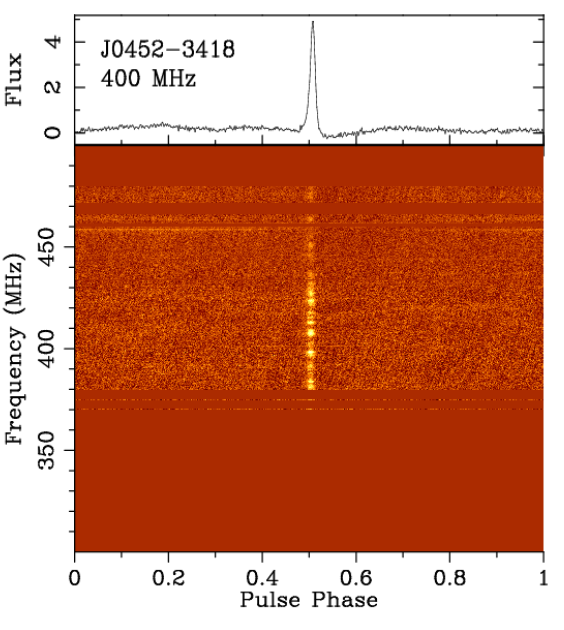}
    \caption{An integrated profile in the top panel and a frequency vs phase plot of a 15\,min snippet of observation GM5 that was not significantly affected by RFI. The flux here is in arbitrary units. }
    \label{fig:gmrtdetection}
\end{figure}

Imaging a later observation (shown in the right panel of Fig.~\ref{fig:psrfiveimage}) revealed a new (and fairly bright) source, denoted as position G, with an estimated flux density of 3.57$\pm0.35$\,mJy (at 400\,MHz), and subsequent observations confirmed this source as the pulsar, with a much-improved detection significance (S/N$\sim$60 in 30 min), as shown in Fig.~\ref{fig:gmrtdetection}. The moderate detection at position D was due to the size of the PA beam being larger than the angular offsets of positions D and G. The pulsar position (i.e., source G) was estimated to be R.A. 04h52m00.7s, and Dec. $-34^\circ 18^{\prime}42.4^{\prime\prime}$.
While the nominal positional uncertainty is $\sim 0.3^{\prime\prime}$, as noted by \citet{psrone} and \citet{smarttwo}, there is likely a systematic offset of ${\sim}3\text{-}4^{\prime\prime}$, likely due to ionospheric refraction (cf. \citealt{swainston2022}). We, therefore, quote a final position as R.A. 04h52m00.7(3)s, and Dec. $-34^\circ 18^{\prime}42(4)^{\prime\prime}$. 

A closer examination suggests a marginal (${\sim}2\sigma$) detection of the pulsar in earlier observations made on 1 May 2023, with an estimated flux density of $0.32 \pm 0.16$\,mJy (at 400 MHz). 
This implies an order-of-magnitude increase in the pulsar's flux density between the two observing epochs (separated by $\sim$2 weeks). Analysis of later observations (GM3 to GM5; see Table 1) confirms the high variability, 
where the pulsar's flux density varied from $\sim$0.7\,mJy (GM4) to $\sim$4.5\,mJy (GM3). 
While this may seem anomalous, especially given the moderate latitude of the pulsar (Galactic latitude, $b = -38.54^{\circ}$), and a distance of $\sim$0.98\,kpc \citep[as per the NE2001 model;][]{ne2001}\footnote{The YMW16 electron density model by \citet{ymw16} yields a larger distance estimate of $\sim$1.34\,kpc.}, and difficult to reconcile solely in terms of refractive scintillation, we note that similarly large flux density variations have been reported for nearby pulsars such as PSRs~B0950+08 and J0437-4715 \citep{2016MNRAS.461..908B,2018ApJS..238....1B}.
Further investigation of this will be reported in a future publication.

\begin{table}[ht]
\centering
    \caption{Summary of pulsar parameters for \psrfive.}
    \begin{threeparttable}
        \bgroup
        \begin{tabular*}{\linewidth}{l@{\extracolsep{\fill}} c}
        \toprule
        
           Timing Parameter & Value \\
           \hline
           Pulsar & \psrfive \\
            Right Ascension (J2000) & 04h52m00.7(3)s\tnote{\textit{a}}\\
           Declination (J2000) & $-34^\circ 18^{\prime}42(4)^{\prime\prime}$\tnote{\textit{a}}\\
           Position epoch (MJD) & 60174\\
           Period (s)  & 1.665118677(1) \\
           Period derivative ($\times10^{-15}$\,s\,s$^{-1}$) & $2.775(3)$\tnote{\textit{a}} \\
           Period epoch (MJD) &  60175 \\
           Dispersion Measure (\dmu) & 19.78(2) \\
           Clock Standard & TT(TAI)\\
           Units & TDB\\
           Time Ephemeris & IF99\\
           Solar System Ephemeris & DE405\\
           \hline
           \multicolumn{2}{c}{Other Measured Parameters} \\
           \hline
           Rotation Measure ($\rmu$) & $-$1.75(8) \\
           $S_{400}$ (mJy) &  2.4$\pm$0.4\tnote{\textit{b}}\\
           \hline
           \multicolumn{2}{c}{Derived Parameters} \\
           \hline
           Magnetic field (G) & $2\times10^{12}$   \\
           Characteristic age (Myr) & 9.5 \\
           $\dot{\rm E}$ (erg\,s$^{-1}$) & $2.3\times10^{31}$\\
           \hline

        \end{tabular*}
        \egroup
        \begin{tablenotes}
        \small
        \item[\textit{a}] These parameters were not derived through a typical timing campaign, see text for more detail.
        \item[\textit{b}] From uGMRT imaging in Band 3 (see Section 2.3 and Fig. 2 for details). The quoted value is the mean flux density from observations GM2 to GM5 (see Table 1 for details). 
        
        \end{tablenotes}
    \end{threeparttable}
    \label{tab:pulsar}
\end{table}


\subsection{Timing}\label{sec:timing}


PSR \psrfive{} was initially timed using the three SMART observations (R04, R09 \& R11), in which the pulsar was detected. These observations spanned $\sim 2$ months and produced an adequate period measurement which allowed for follow-up observations.

We then used two uGMRT observations (GM4 and GM5) to measure a second independent value for period $P$. These observations were intended for nulling and sub-pulse drifting analysis; however, their long durations (2\,hrs each) led to accurate measurements of the period $P$. We also attempted to include uGMRT observations that were used for imaging (GM2 and GM3); however, the observations were contaminated by radio frequency interference (RFI) and some systematic timing errors caused some specific chunks of the observations to misalign in phase. We therefore restricted to observations GM4 and GM5 for the second measurement of $P$. 

As the MWA and uGMRT observations were $\sim 3$ years apart, they could not be readily phase-connected. To obtain a period derivative, $\dot{P}$, we calculated the change in the period between the MWA and uGMRT data sets over the three years. The period derivative and other derived parameters are listed in Table \ref{tab:pulsar}.  

The period derivative determined in this manner is $\dot{P} \sim 2 \times 10^{-15}$\,s/s. With a period of $\sim 1.67$\,s, this $\dot{P}$ measurement places the pulsar slightly to the right of the main island of pulsars in the $P - \dot{P}$ diagram. More precise measurements may be needed to accurately place the pulsar on the $P - \dot{P}$ diagram.
The implied surface magnetic field is $B_{\rm s} = 3.2\times\sqrt{P\dot{P}}$ $\sim 2 \times 10^{12}$\,G and the characteristic age (using $\tau = \frac{P}{2\dot{P}}$) is $\sim9.5$\, Myr. These parameters are fairly typical 
of a `middle-aged' pulsar approaching the death lines.

\section{Analysis and Results \label{sec:analysis}}

\begin{table*}
    \centering
    \caption{Summary of the key parameters measured for \psrfive{} in MWA observations}
    \begin{threeparttable}
    
    \bgroup
    \def\arraystretch{1.5}%
    \begin{tabular*}{\textwidth}{c@{\extracolsep{\fill}} c c c c c c c   }
    \toprule
    
         \multirow{2}{*}{Observation} & DM &RM& $P_3$ & $P_2$ &  $n_{\rm f}$ &$N_{\rm P}$ & Fractional \\
         &\dmu{}&\rmu{}&(Periods) & (deg) &(\%)&(Periods)&Polarisation \\
         \hline
         R04&$19.81\pm0.03$&$-1.75\pm0.08$& $4.89^{+1.60}_{-0.97}$&$6.9^{+9.5}_{-2.5}$&35$\pm8$&$45^{+20}_{-16}$&$<0.6$\\
         R09&$19.85\pm0.07$& -&$4.66^{+1.09}_{-0.74}$&$6.9^{+6.9}_{-2.3}$&32$\pm17$&$41^{+3}_{-2}$&-\\
         R11&$19.72\pm0.07$& -&$4.69^{+2.00}_{-1.02}$&$8^{+22}_{-3.4}$&27$\pm10$&$73^{+62}_{-34}$&-\\
         
         \hline
         \end{tabular*}
        \egroup
        
    \end{threeparttable}
    \label{tab:parameters}
\end{table*}

\subsection{Nulling Analysis}\label{sec:null}

In order to characterise the nulling behaviour of the pulsar, we first calculated the integrated flux density of each pulse by summing over the phase range defined by the on-pulse region. The on-pulse region was defined based on the integrated pulse profile, and a visual inspection for the observation in the form of displayed pulse stacks.  
The time series of the integrated flux estimated for each observation are shown in Figure \ref{fig:nulling}. Further visual inspection of the pulse stack (e.g., see the right panel of Figure \ref{fig:profile+stack}) suggests that the fractions of nulling pulses and non-nulling pulses are roughly equal. The figures also show a quasiperiodic behaviour of nulling.

A commonly adopted method for measuring the nulling fraction of a pulsar is via the Ritchings \citep{Ritchings1976} method. This trials several values of nulling fraction ($n_{\rm f}$)  and chooses one that minimises the expression $\text{ON} - n_{\rm f} \times \text{OFF}$ for intensities $\leq 0$. Here ON and OFF are the on-pulse and off-pulse distributions. The algorithm assumes that all intensities in the on-pulse region are due to nulling. However, as the cut-off for positive and negative intensities is defined by the noise, the measured nulling fraction may differ from its true value, depending on the sensitivity of the telescope. For pulsars with relatively low S/N per pulse, which will have on-pulse histograms close to the off-pulse histograms, 
this may potentially result in the nulling fraction being overestimated, as shown by \citet{Kaplan2018} and \citet{Anumarlapudi2023}.

\begin{figure}
    \centering
    \includegraphics[width=\linewidth]{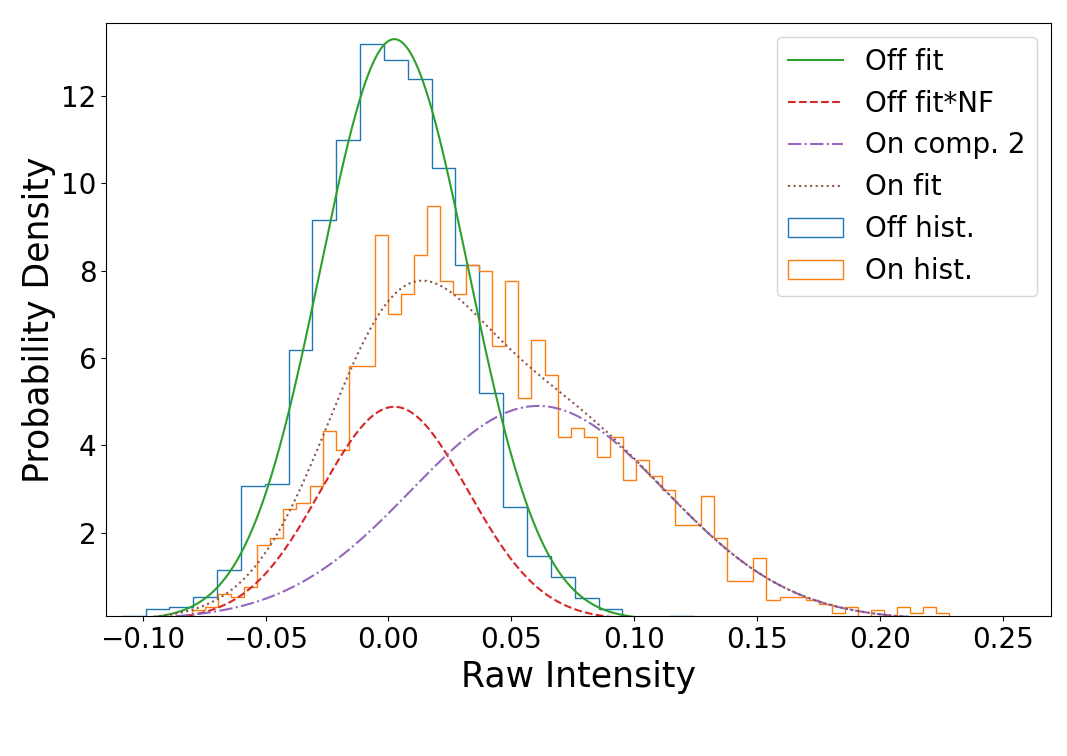}
    \caption{On-pulse and off-pulse histograms (in orange and blue, respectively) from the observation R04. The off-pulse histogram is fit with a simple Gaussian (green solid line), whereas the on-pulse histogram is fit with a multicomponent Gaussian (brown dotted line). The multicomponent Gaussian is composed of a Gaussian that represents real emission (purple dot-dash line), and a nulling Gaussian (red dashed line) that is a scaled version of the off-pulse Gaussian.}
    \label{fig:hist}
\end{figure}

\begin{figure*}
    \centering
    \makebox[1\textwidth][c]{\includegraphics[width=\textwidth]{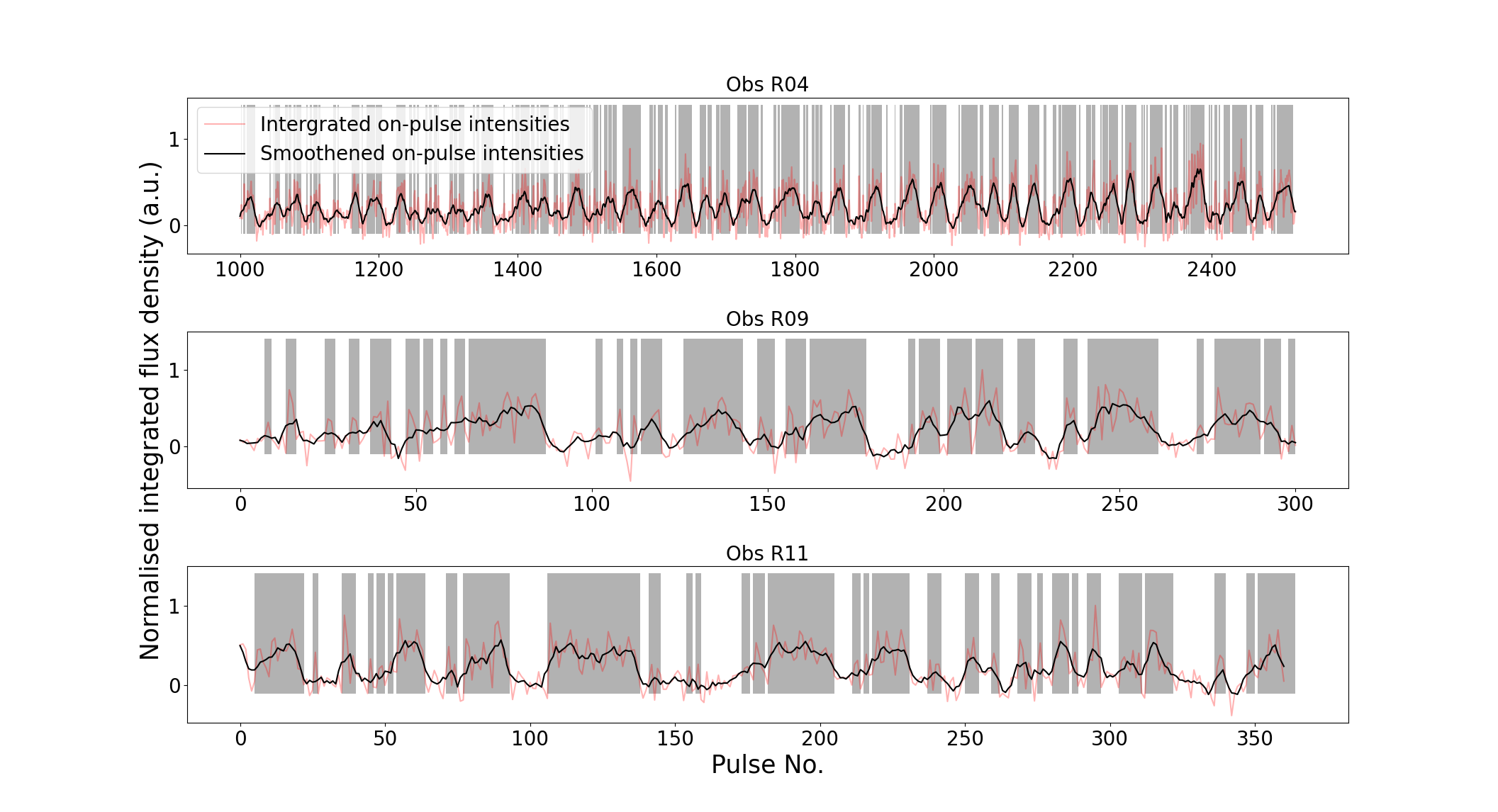}}
    \caption{The three panels show the integrated flux of the on-pulse region (with respect to phase) for each pulse number from SMART observations: R04, R09 and R11. As indicated by the legend, the red lines show the integrated on-pulse intensities and the black is the result of smoothing the data using a Gaussian kernel of the width of the flux in red in order to highlight its quasiperiodic nature. The grey bars indicate emission, i.e. non-nulling regions, based on the data in red. They are present as they help recognise the patterns in the nulls. An intensity of 0.1 was chosen, via visual inspection, as the cut-off for nulling. }
    \label{fig:nulling}
\end{figure*}

As PSR \psrfive{} is only moderately bright in the MWA band, the signal-to-noise ratio (S/N) per pulse is rather low; we therefore used the recently developed method that utilises Gaussian Mixture models, an implementation of which is given in \cite{Kaplan2018}. This relies on having both on-pulse and off-pulse histograms for the pulse energies, just like the Ritchings method. 
The method however utilizes Markov chain Monte Carlo (MCMC) to fit $N$ Gaussians to the on-pulse region, which makes up the nulling and non-nulling distributions, see Figure \ref{fig:hist} for an example of the decomposition. The algorithm then compares the Gaussian fit for the nulling region to the Gaussian for the off-pulse region to estimate the nulling fraction. This method also provides an uncertainty for the measured nulling fraction, which is not readily provided by the Ritchings method. \cite{Kaplan2018}'s implementation was further expanded by \cite{Anumarlapudi2023}. It is publicly available through the \texttt{pulsar\_nulling}\footnote{\url{https://github.com/AkashA98/pulsar_nulling}} software. \cite{Anumarlapudi2023} also included additional models such as Gaussians with exponential tails for pulsars with occasional bright pulses and improved general usability.

Using \texttt{pulsar\_nulling}, we measured nulling fractions for the three observations. We note that \texttt{pulsar\_nulling} expects the on-pulse region as a 2D array, however,  this leads to an overestimation of the nulling fraction. It introduces the noise between pulse components into the on-pulse histogram which leads to an overestimated nulling fraction measurement. We instead defined the on-pulse region as a 1D array that had been averaged over the pulse phase. We measured the nulling fraction to be $35\pm8\%$ for R04, $32\pm17\%$ for R09, and $27\pm10\%$ for R11 (see Table \ref{tab:parameters}).


In estimating the nulling fractions, averaging the on-pulse region turns out to be an appropriate approach for this pulsar. When the full on-pulse region is used, the algorithm  (\citet{Anumarlapudi2023}) generates pixel-dependent on-pulse histograms, which, in the case of low S/N data, can result in less reliable results. Our estimated nulling fractions are $55\pm2\%$ for R04, $69\pm6\%$ for R09, and $67\pm4\%$ for R11, and our method is justifiable given that the on-pulse region is much wider than a sub-pulse. However, as noise is also added, this may result in some overestimation of nulling parameters.



The variation between the three MWA observations is likely to do with the fact that each observation only contains a relatively small number of null sequences. Accounting for the length of the observations, a weighted average nulling fraction is $ 34\pm6\%$. Given the limited data sets, this weighted average does not represent the mean of the true distribution of nulling fractions from the pulsar, but just the mean from our observations.

Although the Gaussian mixture model makes a more robust
tool for analysis, it can still overestimate the nulling fraction if the chosen on-pulse window is broader than the intrinsic pulse width. This can become an issue when dealing with pulsars such as giant pulse emitters that have normal pulses and giant pulses offset in phase (as seen in PSR~J1047$-$6709 by \citealt{2021MNRAS.501.3900S}). As the method creates a histogram of the pulse energy distributions, a wider-than-necessary on-pulse region may introduce extra noise. This may increase the component of the on-pulse region histogram that is in line with the off-pulse region histogram, thereby increasing the measured nulling fraction. This can also affect the Ritchings method in similar ways, as that also relies on a histogram of pulse energy distributions to measure the nulling fraction.

\subsubsection{Quasiperiodicity}

To measure the quasiperiodicity of nulling, we produced an FFT of the pulse energies obtained by summing over the on-pulse region.
We fit a normal distribution to the log of the frequencies as a means to measure a mean period. The mean value of the distribution is interpreted as the nulling frequency whose inverse (the nulling period) is listed in Table \ref{tab:parameters}. The uncertainty was measured as the variance of the fit of the mean in the frequency domain and converted to the time domain; similar to the $P_3$ values. The average nulling period across R04, R09 and R11, weighted by the number of pulses, is 45$^{+31}_{-13}$\,$P$ (where $P$ is the pulsar period). We also summed the FFTs of all three observations, as seen in the left panel of Figure \ref{fig:2dfs}, to observe any long-term behaviour between observations. We reprocessed the observations to have the same number of bins (512) for this summation. We also only summed 301 sub-integrations of an observation at a time to ensure they all have the same Fourier bins. We again fit a Gaussian distribution to the logarithm of the combined FFT and find a $P_3$ value of $42^{+1.5}_{-1.3}\,P$.

\begin{figure*}[ht]
    \centering
    \includegraphics[width=0.9\linewidth]{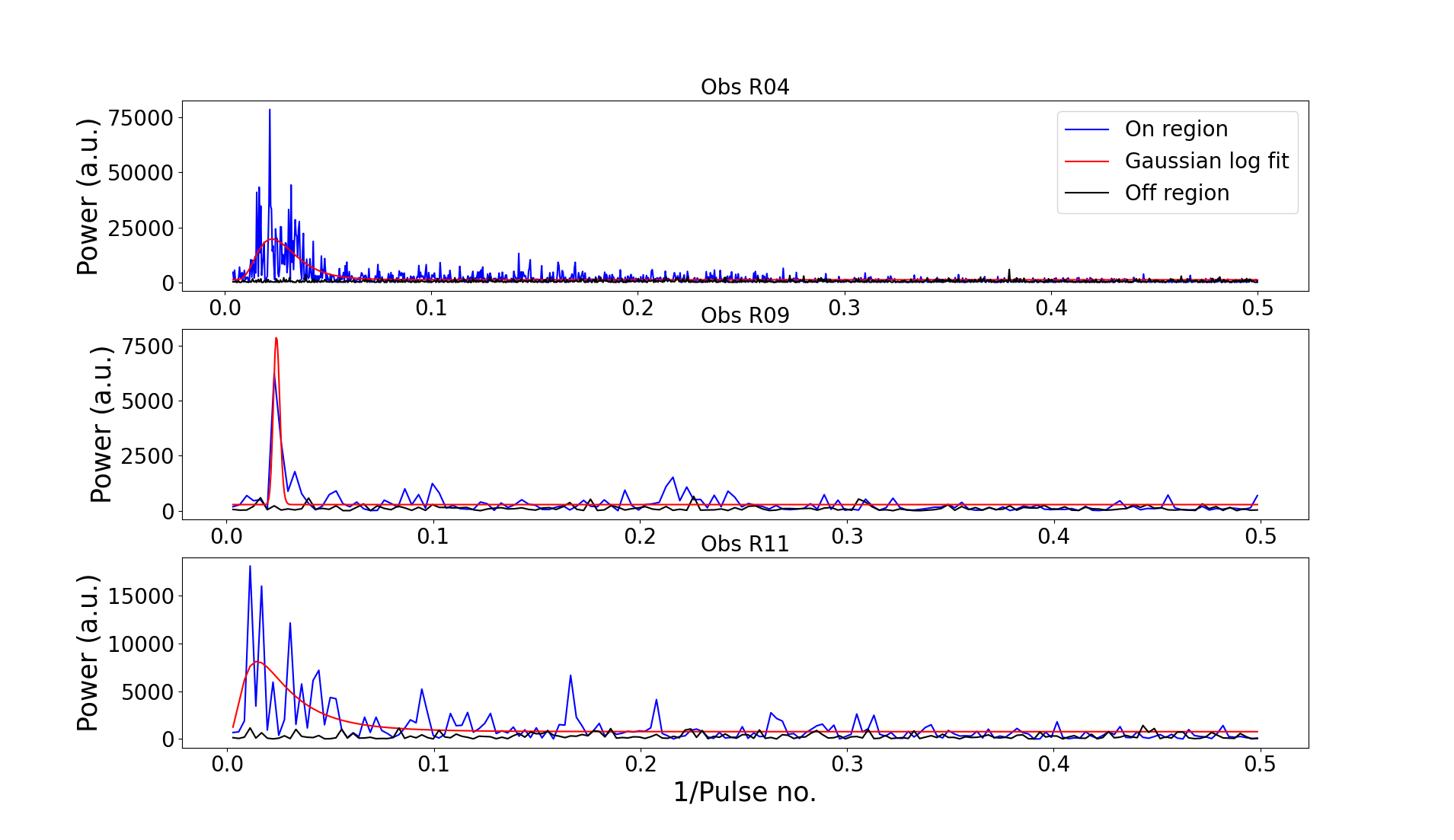}
    \caption{An FFT of the on-pulse region and the off-pulse region of \psrfive{} in observations R04, R09 and R11, shown in blue and black, respectively. A normal distribution is fit to the log of the FFT, specifically for the peak in the on-pulse region that is caused by the quasiperiodicity in the nulling, shown in red.  }
    \label{fig:r09fft}
\end{figure*}

\subsection{Sub-pulse drifting}\label{sec:subdrift}

Upon visual inspection, most drift sequences for PSR \psrfive{} appear to be organised. The drift sequences tend to contain on average four drift bands but can range anywhere from one to eight bands. Overall the pulsar appears to have a stable drift rate as seen in Figure \ref{fig:profile+stack}. However, some drift sequences appear less organised and fairly diffuse, which may have different characteristics that are hard to discern due to low S/N. The drift sequences for most pulsars are observed to start and stop rather abruptly, but in the case of PSR \psrfive, there appears to 
be diffuse emission near the ends of drift sequences; however, this could also be an artefact resulting from low S/N in our observations.

We produced a 2-dimensional fluctuation spectrum \citep[2DFS;][essentially a 2-dimensional FFT of the pulse number against the pulse phase plane]{2002A&A...393..733E} of the on-pulse region. This is the most commonly used method to measure $P_2$ and $P_3$ values. As some pulsars show either multiple values of $P_3$ (in the case of multiple drift modes, e.g. PSR~J0034$-$0721; \citealt{McSweeney2017}) or a slowly evolving $P_3$ (e.g. PSR~J0026$-$1955; \citealt{Mcsweeney2022, Janagal2022}). We report a $P_3$ value for \psrfive{} for each observation where it is measurable (see Table \ref{tab:parameters}). For completeness, the table also includes $P_2$ values for each observation.

\begin{figure}[ht]
    \centering
    \makebox[\linewidth][c]{\includegraphics[width=\linewidth]{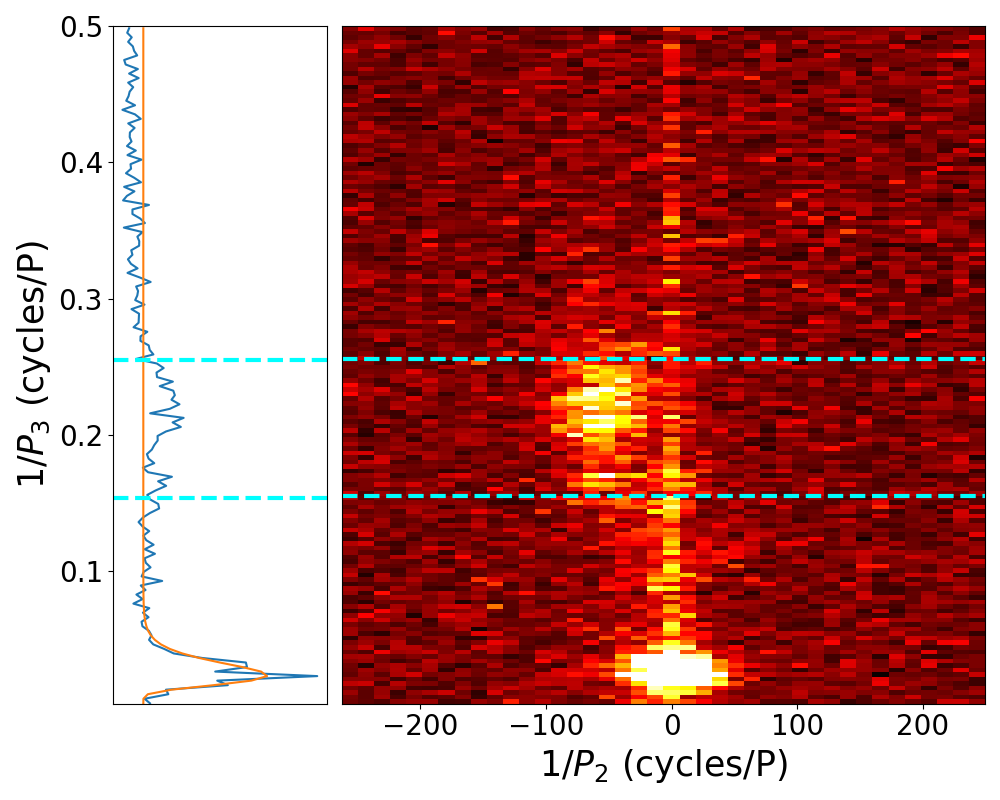}}
    \caption{A 2DFS of the on-pulse region in R04, R09 and R11 summed together (right panel), with the 1D cutout of each row in the left panel. The cyan lines highlight the sub-pulse drifting in both plots. The bottom-most feature in both panels is due to the quasiperiodicity in the nulling. }
    \label{fig:2dfs}
\end{figure}

We produced 2DFS for each of the three observations (R04, R09 and R11); however, due to low S/N, the peaks are not very obvious visually. For clarity, we therefore summed the 2DFS from the three SMART observations, and this is shown in Figure \ref{fig:2dfs}. We identify peaks in this spectrum that refer to the $P_3$ values, which are outlined with cyan lines in Figure \ref{fig:2dfs}. We also show the one-dimensional FFT in the left panel of Figure \ref{fig:2dfs}, which reveals a broad peak and a smaller separate peak below 0.2. The broad features with multiple peaks were seen consistently in the three MWA observations. Using only the summed 2DFS, we measured features associated with the sub-pulse drifting, and these are located at $1/P_2 = f_2 = 50.4\pm39.6$\,cycles/$P$ ($P_2 = 7.1^{+26.3}_{-3.1}$ degrees) and $1/P_3 = f_3 = 0.21\pm0.05$ rotations ($P_3 = 4.8^{+1.5}_{-0.9}$ pulses). The uncertainty for these values is asymmetric as the uncertainty was determined in Fourier space and converted to temporal.

\subsection{Polarisation}

Propagation through the magneto-ionised interstellar medium (ISM) causes the plane of linear polarisation to rotate as a function of wavelength squared, at a rate proportional to the rotation measure (RM; the integrated magnetic field strength along the line of sight, weighted by the electron density).
This effect, known as Faraday rotation, can cause the frequency-averaged signal to be depolarised, even for relatively low RM values.
The polarisation can be recovered by removing the Faraday rotation using a measured RM.
We measured the RM using RM-synthesis \citep{Burn1966,Brentjens2005}, a robust technique that isolates the astrophysical RM from instrumental polarisation with weak or no spectral dependence (e.g., polarisation leakage).
Specifically, we used a Python implementation of RM-synthesis made available by \citet{Heald2009} via a public repository\footnote{\url{https://github.com/gheald/RMtoolkit}}.

For the polarisation analysis, we have only used the R04 observation as it was the longest and most sensitive of all three MWA observations. We note that absolute calibration of the MWA for reliable polarimetry is still a work in progress \citep{2019PASA...36...25X} and as a result, our polarisation profile is accurate only as a first-order estimate.
We measure a total RM of $-2.21\pm0.04\,\rmu$, and after accounting for an ionospheric contribution of $-0.46\,\rmu$, this yields an ionosphere-corrected RM of $-1.75\pm0.08\,\rmu$, which is similar to that reported for the nearby pulsar J0448$-$2749 that has a catalogue RM of $-0.7\pm5.4\,\rmu$ \citep{Han2018}. However, the RM comparison of the two pulsars is 
not so meaningful as the uncertainty for J0448$-$2749's RM is very high compared to the very precise measurement for \psrfive{} obtained via low-frequency observations. We also noted that the RM of PSR~\psrfive{} is rather low compared to those reported for previous SMART discoveries PSRs \psrone{} and \psrtwo{}, both of which are located at much higher Galactic latitudes \citep{psrone,smartone}.

\begin{figure}[ht]
     \centering
     \includegraphics[width=\linewidth]{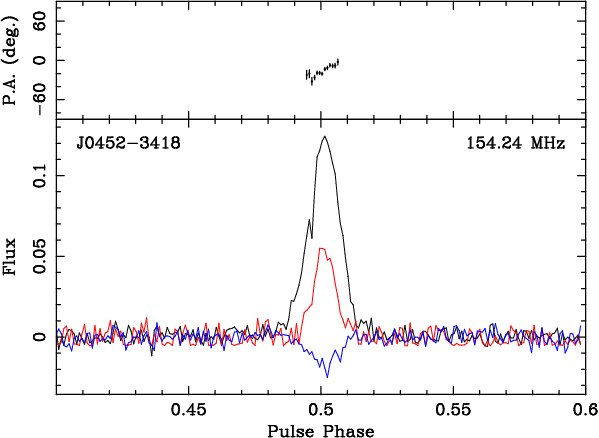}
     \caption{The bottom panel shows the integrated profile of \psrfive{} from MWA observation R04. The total intensity is shown in black, the linear polarisation in red, and the circular polarisation in blue. The top panel shows the polarisation angle of linear polarisation. The flux here is in arbitrary units.}
     \label{fig:pol}
 \end{figure}

The Faraday rotation was removed using the \texttt{pam} routine from \code{psrchive} using the observed RM of $-2.21\pm0.04\,\rmu$.
The resulting polarisation profile (see Figure \ref{fig:pol}) shows moderate levels of linear polarisation and low circular polarisation, with an overall polarisation fraction $<0.6$.

The sweep in the position angle of linear polarisation as a function of the pulse phase is consistent with that seen for most pulsars.
We estimate the gradient of the polarisation angle curve to be $14.1\pm1.6\,\mathrm{deg}\,\mathrm{deg}^{-1}$. We also attempted to fit the rotating vector model (RVM; \citealt{RVM1969}) to the polarisation angle sweep; however, due to the narrowness of the pulse, the fit was rather unconstrained.
Using the measured RM and DM, we calculate the average magnetic field strength parallel to the line of sight as $\langle B_\parallel\rangle=1.232(\mathrm{RM}/\mathrm{DM})\,\mu\mathrm{G}$.
The measured $\langle B_\parallel\rangle$ is $-0.11\pm0.08\,\mu\mathrm{G}$, which is consistent with a value of $-0.033\pm5.4\,\mu\mathrm{G}$ reported for J0448--2749 \citep[using the DM from][]{Hobbs2004}.

\section{Discussion}\label{sec: disc}

PSR~\psrfive{} has a narrow profile though there is a hint of two components at low frequencies (155\,MHz), based on Figure \ref{fig:pol}. As seen from this figure, our observations show a modest amount of linear polarisation and a small amount of circular polarisation. Based on these, we would hence classify the observed emission as coming from the conal beam of the pulsar \citep{Rankin1983}. Furthermore, the pulsar also has a relatively long period and it shows sub-pulse drifting; these are typical traits of pulsars that exhibit a conal emission \citep{Basu2016}.


\subsection{Sub-pulse drifting}

PSR \psrfive{} appears to be a predominantly stable drifter with fairly organised drift sequences and a singular drift mode punctuated by quasiperiodic nulls. However, it is unclear if some of the fainter drift sequences have slightly different drift rates.
Further, there appears to be some faint emission near the start and end of some drift sequences as they appear to transition to nulling episodes. 
Most nulling pulsars tend to show a very sharp transition between the nulling and burst states, suggesting that the transition between the two states occurs on a time scale much smaller than one rotation. Higher S/N observations would be needed to ascertain 
if this is also the case for PSR \psrfive{}, or if the apparent faint emission that occurs at the drift sequence boundaries is indicative of a more gradual transition.
As well known, some pulsars tend to display a variable drifting behaviour, marked by multiple distinct drift modes, while others show a generally consistent drift 
pattern throughout. 
In the case of the former, the fluctuation spectra show multiple sharp $P_3$ peaks, whereas somewhat broader $P_3$ peaks are seen in the case of those with regular drift sequences (see Appendix of \citealt{Weltevrede2006} for examples). Our observations show that PSR \psrfive{} shows broad and diffuse peaks in the 2DFS, as shown in Figure \ref{fig:2dfs}. The broadness may be due to multiple intrinsic values of $P_3$, or a slowly changing $P_3$;  this is possibly due to the diffuse emission near the ends of drift sequences. 

To test the stability of $P_3$, we attempted to fit a quadratic function to the drift rate for each drift sequence in the observations, as was done for PSR \psrtwo{} by \citet{Mcsweeney2022}. However, the short durations of the drift sequences resulted in fits being only marginal significance. Furthermore, the fits were very sensitive to the weaker, possibly less organized single pulses near the drift sequence boundaries. We therefore were unable to reject the null hypothesis that $P_3$ is unchanging.

PSR J1631+1252 shows broad features in its 2DFS \citep{Wen2022}; however, it also exhibits a bimodal distribution of $P_2$, which implies multiple emission modes. Our analysis does not suggest a bimodal distribution in $P_2$ for \psrfive, though this may also be due to our limited data sets.


Importantly, PSR \psrfive{} fits comfortably within the population of sub-pulse drifting pulsars that also show quasiperiodic nulling. In Table \ref{tab:quasipulsars}, we compare the nulling/drifting properties of this pulsar with those of others within this subclass, collated from various studies published over the past $\sim$20 years. This population tends to have rotation periods around $\sim$1-2\,s, nulling fractions in the range $\sim$5-50\%, and nulling periods ranging from a few tens of pulses to a few hundred. PSR \psrfive's rotation period and nulling period are quite close to the respective median values, however, its nulling fraction is above average.

Within this particular subset, there appears to be no obvious correlation between sub-pulse properties (e.g., $P_3$ and whether there are multiple drift modes) and nulling properties (e.g., nulling fraction, nulling period).
There is also no obvious difference between the sub-pulse drifting properties of this subset compared to those that do not exhibit quasi-periodic nulling \citep[e.g. the sample in][]{Weltevrede2006}, which suggests that the quasi-periodicity in nulling may not be necessarily connected to sub-pulse drifting. 
Notwithstanding this, there has been some published work that attempts to explain quasi-periodic nulling as a byproduct of the carousel model. Below, we discuss whether this model is applicable for PSR \psrfive{}.

\begin{table*}[]
\centering
\caption{A list of sub-pulse drifting pulsars reported to show quasiperiodic nulling in \cite{Basu2020}, and their relevant parameters. Modes without a clear sub-pulse drifting signature are indicated with a hyphen in the $P_3$ column.}
    \begin{threeparttable}
    \bgroup
    \def\arraystretch{1.5}%
    \begin{tabular*}{\textwidth}{l@{\extracolsep{\fill}}cccccl}
\hline
Name & \multicolumn{1}{c}{Period } & \multicolumn{1}{c}{Nulling Period } & \multicolumn{1}{c}{Nulling Fraction } & \multicolumn{1}{c}{$P_3$ } & Modes & References \\ 
& \multicolumn{1}{c}{ (s)} & \multicolumn{1}{c}{ ($P$)} & \multicolumn{1}{c}{ (\%)} & \multicolumn{1}{c}{($P$)} &  &  \\ \hline
J0034-0721   & 0.94 & 75$\pm$14  & 44$\pm$1 & 13$\pm0.8$, 7$\pm0.2$, 5$\pm0.3$    & 3 & (2, 6, 13)   \\
J0304+1932   & 1.38  & 103$\pm$34  & 14$\pm$4  & 6$\pm1.7$    & 1 & (3, 11)     \\
\psrfive{}   & 1.67 & 42$^{+1.5}_{-1.3}$ & 34$\pm6$ & $4.8^{+1.5}_{-0.9}$ & 1 & This work \\
J0837+0610   & 1.27   & 16$\pm$4  & 9$\pm$1   & 2.2$\pm0.03$   & 1  & (3, 10)      \\
J0934-5249   & 1.44   & 35$\pm$19  & 5$\pm 3$    & 4$\pm$0.2  & 1 & (1, 7, 10)     \\
J1239+2453   & 1.38    & 26$\pm$5   & 7$\pm 3$  & 3$\pm0.1$  & 1 & (1, 7, 10)    \\
J1727-2739 & 1.29   & 206$\pm$33  & 52$\pm 3$   &10$\pm1.6$, 5$\pm0.9$, -   & 3  & (2, 8, 12)    \\
J1741-0840   & 2.04 & 34$\pm$8   & 16$\pm 1.7$   & 5$\pm0.6$   & 1  & (2, 11)     \\
J1822-2256   & 1.87 & 134$\pm$33  & 5$\pm$0.9   & 17.9$\pm0.02$, 6$\pm0.2$, 8$\pm0.2$, 14.1$\pm0.02$  & 4 & (4, 2, 14)    \\
J1921+1948   & 0.82   & 85$\pm$14   & $\sim$2 & 6$\pm0.3$, 4$\pm0.1$, 2.4$\pm0.04$, -  & 4 & (3, 2, 10)    \\
J1946+1805   & 0.44   & 600$\pm$52   & 30$\pm1.4$   & 13.8$\pm0.07$, 6$\pm1.7$, -, -  & 4 & (2, 10)    \\
J2006-0807   & 0.58  & 41$\pm$4 & 16$\pm 1$  & 15$\pm2.5$   & 1  & (5, 2, 10)    \\
J2048-1616   & 1.96  & 51$\pm$20 & 8$\pm1.4$  & 3.2$\pm0.03$ & 1  & (2, 10)     \\
J2305+3100   & 1.57 & 43$\pm$8 & 5$\pm 0.5$  & 2.1$\pm$0.05, ${\sim} 3$   & 2 & (3, 2, 10)    \\
J2313+4253   & 0.34   & 32$\pm$11   & 4$\pm 0.5$   & 2$\pm$0.1    & 1  & (2, 10)    \\
J2321+6024   & 2.25  & 58$\pm$17  & 36$\pm1.2$  & 8$\pm$1, 4$\pm$1, 3$\pm0.5$  & 3  & (1, 9, 10)    \\
\hline
\end{tabular*}
\egroup
        \begin{tablenotes}
        \small
        \item[] \textbf{Note:} The periods are obtained from the \href{https://www.atnf.csiro.au/research/pulsar/psrcat/}{ATNF Pulsar Catalogue} \citep{ATNF}
        \item[] \textbf{References:} 
        1 - \cite{Basu2020}; 
        2 - \cite{Basu2017}; 
        3 - \cite{Herfindal2009};
        4 - \cite{basuMitra2018};
        5 - \cite{Basu2019};
        6 - \cite{Gajjar2017};
        7 - \cite{Naidu2017};
        8 - \cite{Wang2007};
        9 - \cite{Rahaman2021};
        10 - \cite{2019MNRAS.482.3757B};
        11 - \cite{Basu2016};
        12 - \cite{Wen2016};
        13 - \cite{McSweeney2017};
        14 - \cite{Janagal2022}

        \end{tablenotes}
\end{threeparttable}
\label{tab:quasipulsars}
\end{table*}
\break
\subsection{Nulling phenomenon}

Nulling is a poorly understood phenomenon of pulsar emission. In the context of sub-pulse drifting, nulling is generally observed as the transition state between neighbouring sparks \citep{Ruderman1975}. The work of \cite{Wang2007} suggests that nulling is often observed when the magnetosphere of the pulsar is going through large-scale changes and that nulling refers to a transition to different emission modes. This is clearly observed with PSR J1701-3726 for example, where its two emission modes are separated by a null \citep{Wang2007}. Furthermore, \cite{2020Wen_modechange} analyzed the mode changing of J1326-6700 and found quasiperiodicity between the transitions. Hence, pulsars that show quasi-periodic nulling may provide a potential link connecting these phenomena. However, based on our current observations, PSR~\psrfive{} does not seem to be displaying multiple emission modes. 
We further note that there are indeed multiple pulsars (Table \ref{tab:quasipulsars}) that show sub-pulse drifting and quasi-periodicity in nulling but exhibit only a single emission mode.


\subsection{Quasiperiodicity in Nulling}

The quasi-periodicity in the nulling behaviour of PSR \psrfive{} is evident in the top panel of Figure \ref{fig:nulling}, which shows the 
integrated flux density estimated 
in the on-pulses region. However, for observations R09 and R11, the quasiperiodicity 
is somewhat less evident. It is visually noticeable in pulses ${\sim} 120$ to $300$ for R09 and in pulses ${\sim} 10$ to $100$ for R11; however, for other pulses, the null and burst durations are either longer or variable. Regardless, the FFTs of the on-pulse region of these observations, as shown in Figure \ref{fig:r09fft}, show consistent results of quasi-periodic nulling occurring every ${\sim}48$ pulses. 
 
\cite{Herfindal2007, Herfindal2009} have attempted to explain the nature of nulling and its quasi-periodic behaviour within the context of the popular carousel model \citep{Ruderman1975}. The carousel model describes pulsar emission in terms of beamlets (a ring of smaller emission beams) that rotate around the magnetic axis like a carousel. In this framework, short-duration nulling (lasting for only a few pulses) may be explained as the precession of the beamlets, with the nulls appearing when the line of sight passes between beamlets \citep{Basu2017}. The nulls could also be missing or empty beamlets themselves, also referred to as `pseudo-nulls' \citep{Herfindal2009}. Therefore, the rotation of the beamlets can introduce periodicity in the nulling as we regularly observe the gaps. However, when pulsars switch to long-duration nulls, the explanation is no longer consistent \citep{Basu2017}.

It is conceivable that the interpretation by \cite{Herfindal2007, Herfindal2009} involving the mechanics for quasi-periodic nulling in conjunction with the spark carousel model can accommodate the observed behaviour of PSR \psrfive{}. 
Specifically, given the clear periodicity seen in some observations, and the seemingly stable sub-pulse drifting behaviour, the nulling in \psrfive{} could be interpreted as ``pseudo nulls", or missing beamlets in the carousel. Given that the nulling period is $\sim 48$ pulses, and the average drift sequence has 4 drift bands, we can interpret there may be 8 - 48 beamlets. Assuming each drift band is a single beamlet slowly moving out of our line of sight, there can be $\sim 4$ consecutive, emitting beamlets and $\sim 4$ consecutive nulling, or pseudo-null, beamlets. Alternatively, the carousel rotation could be slightly faster than the pulsar rotation such that in each pulsar period, we see the next consecutive beamlet. In this scenario, there can be $\sim 24$ emitting beamlets and $\sim 24$ pseudo-null beamlets.

Although PSR \psrfive{} shows clear evidence of quasiperiodicity, there are instances of some departure from this general trend. 
For example, Figure \ref{fig:nulling} shows pulses 0 - 100 in R09 and pulses 100 - 361 in R11, where the pulses appear to have no set nulling period. Additionally, the pulsar occasionally shows isolated drift bands and null-interrupted drift bands, which add a further layer of complexity when applying \cite{Herfindal2007, Herfindal2009}'s model.

Interpretation in terms of pseudo nulls (missing or blank beamlets) seems a plausible explanation 
in cases where nulling appears to be quasi-periodic. However, most nulling and sub-pulse drifting pulsars have not been reported to show quasi-periodic nulling. Additionally, only $\sim 15$ (quasi)periodically nulling pulsars have also been sub-pulse drifters \citep{Basu2020}. In cases where the pulsar is (quasi)periodically nulling, the behaviour appears to be dominant but not consistent. Therefore, it is possible that nulling is independent of sub-pulse drifting and its quasi-periodic nature is due to other factors.

Aside from the above, there have been few other attempts to explain a quasi-periodic nulling. The only one in the published literature that 
we are aware of is by \cite{Caputo2023} who explains periodic nulling and the pulsar's radio emission in terms of axion clouds. In their theory, light axions can occasionally screen the emission region in the pulsar magnetosphere, producing quasi-periodic nulling and other intrinsic pulse variability. However, their models only apply to specific parameter spaces and require testing on multiple pulsars. Clearly, newer models may be needed to comprehensively account for the observed quasi-periodic nulling and sub-pulse drifting behaviour in PSR \psrfive{}.

\subsubsection{Pulsar nulling and death lines}

\begin{figure}
    \centering
    \includegraphics[width=1\linewidth]{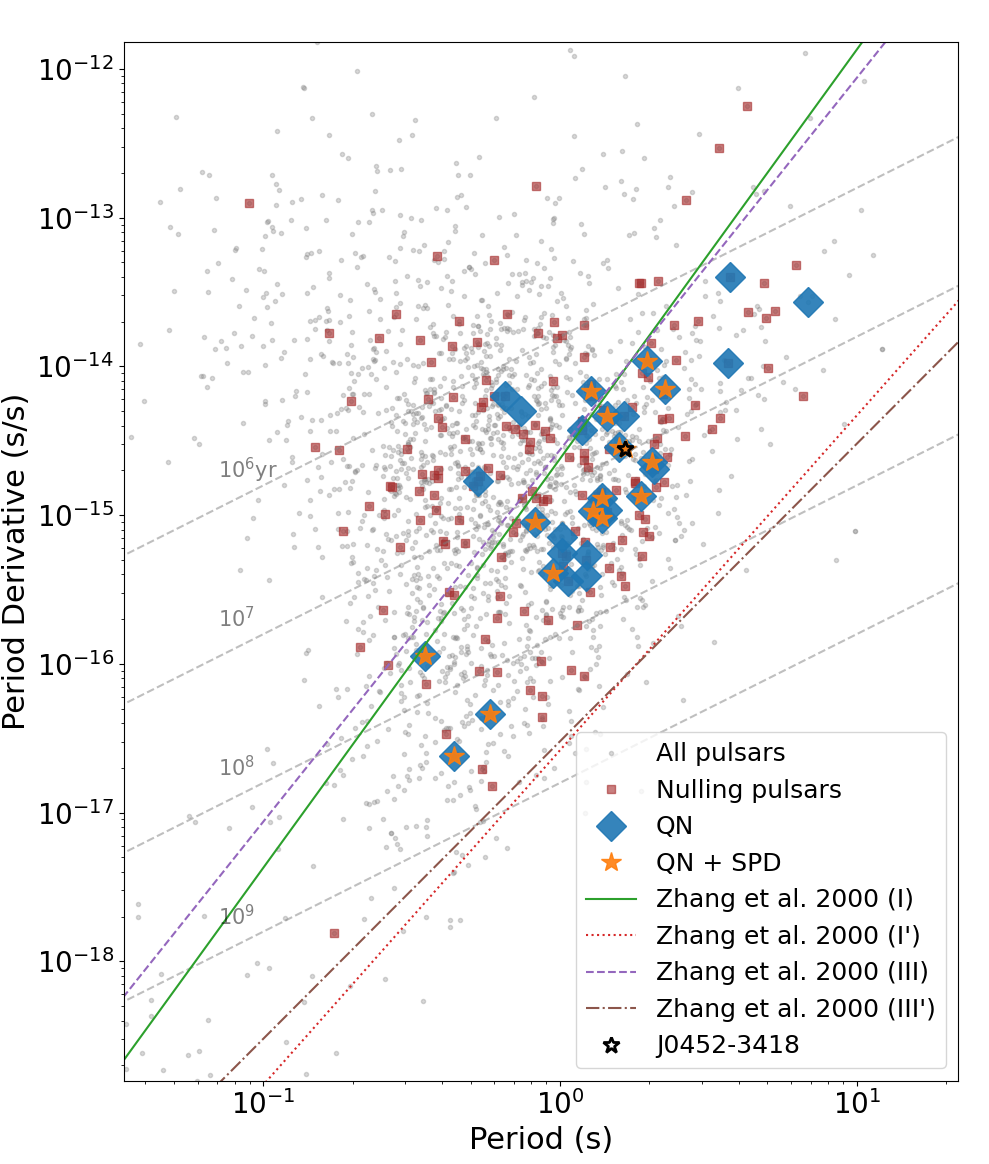}
    \caption{A Period vs Period derivative diagram showing pulsars of different types. All pulsars are shown in grey, all nulling pulsars are shown in brown, QN or quasiperiodically nulling pulsars are shown as blue diamonds and QN + SPD (quasiperiodically and sub-pulse drifting pulsars ) are shown as orange stars. Various death lines from \cite{Zhang2000} are also shown. The grey lines indicated the characteristic age of the pulsars. The quasiperiodically nulling pulsars used in the diagram were derived from \cite{Basu2020,Anumarlapudi2023}.}
    \label{fig:quasippdot}
\end{figure}

In the $P - \dot{ P}$ diagram shown in Figure \ref{fig:quasippdot} we have highlighted the currently known population of nulling pulsars;  the three interesting sub-classes -- {\it viz.} the population that shows nulling; the sub-class that shows evidence for quasiperiodically nulling; and the smaller subset that shows both quasi-periodic nulling as well as sub-pulse drifting -- are highlighted, in an effort to extract some useful clues.
With the caveat that this is a representation of the current population of nulling pulsars, it is intriguing that the sub-population of pulsars that show quasi-periodic in nulling is primarily located in and around the vicinity of the two pairs of death lines (often referred to as `the death valley'). In particular,  the sub-population that shows both quasi-periodic nulling and sub-pulse drifting has a comparatively smaller spread near the death lines, with a majority of them lying well inside the region of the pairs of death lines. Most of these are essentially either fairly older or middle-aged pulsars, with characteristic ages in the range of $\sim$1 to 100 Myr. Interestingly, they also suggest a gradient that is quite similar to those of the death lines. This may suggest that the mechanism responsible for the quasi-periodic nature in nulling may be possibly linked to the proposed death mechanisms for pulsars \citep[c.f.][]{Ruderman1975, 1993ApJ...402..264C, Zhang2000}. It is quite possible that the initial death line conditions do not result in a cessation of emission but rather quasi-periodic fluctuations in the emission mechanism, which is observed as a quasiperiodicity nulling.

Another notable hint from Figure \ref{fig:quasippdot} is that the nulling population may be divided into two separate sub-classes (i.e. quasi-periodic versus non-quasi-periodic nulling), which may be linked to the characteristic age or spin period of pulsars. However, it is important to note that there hasn't been a systematic investigation for a quasiperiodicity in nulling using a large sample of pulsars, and therefore we may not preclude the possibility that many of the pulsars marked in brown may also show quasi-periodic nulling. Such a study can therefore help us further explore the nulling phenomenon in pulsars. Based on Figure \ref{fig:quasippdot}, it appears that longer-period pulsars ($P\gtrsim1$\,s) may be more likely to show this feature. This would hence benefit from a population study, aimed at analysing the presence of quasi-periodic nulling in a sample of pulsars. With the updated census of low-frequency detections from the ongoing SMART pulsar survey \citep{smarttwo}, there is a sizeable sample of pulsars in the southern sky, data for which can be meaningfully utilised for such an analysis. The related analysis is currently in progress (Grover et al. in prep.) and will be reported in a forthcoming publication.

\subsection{Detection prospects of long-period nulling pulsars with future surveys}

The current population of pulsars that shows quasi-periodic nulling is $<1\%$, which makes it difficult to perform any meaningful population analysis. 
As many of these pulsars tend to have longer periods and/or substantial nulling fractions (and null durations), they can be hard to detect using standard pulsar searching methods. 
Until recently, the vast majority of all-sky pulsar surveys (with the exception of SMART and LOTAAS; \citealt{2019A&A...626A.104S}) employed fairly short dwell times of $\sim$2-5 minutes for large parts of the sky, which is rather short to uncover a large population of long-period nulling pulsars. 
This is also the case with FFT-based search algorithms, which have been shown to be less sensitive towards such objects \citep{Kondratiev2009,Morello2020}. 

To detect more such objects, surveys with longer observing duration and employing search software sensitive to long periods are essential. SMART and LOTAAS (the LOFAR pulsar survey) are examples of such surveys as they both have $\sim$1 hour dwell times, which makes them particularly appealing to search for such objects. SMART has already detected multiple long-period pulsars, of which both PSR \psrfive{} and \psrfour{} show both nulling and sub-pulse drifting. 
Furthermore, other instruments such as CHIME/Pulsar \citep{2021ApJS..255....5C} have collected long-duration or regular cadence data, which can be used to search for such long-period nulling pulsars \citep[e.g.][]{2020ApJ...903...81N}. These surveys can therefore eventually yield an improved sample of more objects like \psrfive{} that exhibit both sub-pulse drifting and quasi-periodic nulling behaviour.
In a forthcoming paper (Grover et al., submitted), we have explored the combined use of FFT, fast folding algorithm and single pulse search (the latter two of which are known to be sensitive towards long-period objects) and how effective use of these different search methods can help increase the sensitivity to long-period nulling pulsars.

\section{Conclusion}\label{sec:con}

Nulling is a poorly understood pulsar emission phenomenon
which has been often linked to pulsar characteristic age and putative death. An uncommonly reported characteristic of nulling is its quasi-periodic nature, which is now seen 
in $\sim$10\% of the nulling population. Nulling is often observed alongside sub-pulse drifting, 
a popular interpretation for which involves rotating `beamlets' arranged in the
form of a carousel configuration. 

In this paper, we report the discovery and follow-up of PSR~\psrfive, a long-period ($P=1.67$\,s) pulsar that shows both quasi-periodic nulling and sub-pulse drifting. It was discovered in the SMART pulsar survey that is currently in progress at the MWA. The pulsar was followed up with the uGMRT, which enabled a precise localisation via high-resolution imaging in the Band 3 (300-500 MHz) range. 
The pulsar's spin parameters were measured using a modest number of observations made using these two telescopes.

For the nulling analysis, we used three specific observations from the SMART survey campaigns. Using a Gaussian mixture model implementation in \texttt{pulsar\_nulling}, we measured the nulling fraction of the pulsar, which is $ 34\%$ in the MWA's frequency band (140-170 MHz). We used the FFT method to measure the quasiperiodicity 
in nulling, which revealed clear peaks, corresponding to a nulling period of ${\sim} 42$ pulses. 
While there are some instances of observations where the observed nulling behaviour does not conform to any clear periodicity, overall there is evidence for a quasiperiodicity in our observations. 

For the sub-pulse drifting analysis, we employed the 2DFS method, which was applied to all three SMART observations; this yielded $P_3 \approx 4.8$ pulses and $P_2 \approx 7$\,degrees per pulse. 
Although, on average, the drift sequence tends to have 4 drift bands, there are some instances of isolated drift bands as well in some of our observations. 
Further, the observations reveal the beginning and end of some drift sequences to be either dim or interrupted by nulling. These factors tend to complicate attempts to explain the observations in terms of a carousel model and the possibility of quasi-periodic nulling being pseudo nulls.

\begin{acknowledgments}

This scientific work uses data obtained from \textit{Inyarrimanha Ilgari Bundara}, the CSIRO Murchison Radio-astronomy Observatory. We acknowledge the Wajarri Yamaji People as the Traditional Owners and native title holders of the Observatory site. Establishment of CSIRO's Murchison Radio-astronomy Observatory is an initiative of the Australian Government, with support from the Government of Western Australia and the Science and Industry Endowment Fund. Support for the operation of the MWA is provided by the Australian Government (NCRIS), under a contract to Curtin University administered by Astronomy Australia Limited. This work was supported by resources provided by the Pawsey Supercomputing Research Centre with funding from the Australian Government and the Government of Western Australia, as well as by resources awarded under Astronomy Australia Ltd's ASTAC merit allocation scheme on the OzSTAR national facility at the Swinburne University of Technology. The OzSTAR program receives funding in part from the Astronomy National Collaborative Research Infrastructure Strategy (NCRIS) allocation provided by the Australian Government.
G.G. is supported by an Australian Government Research Training Program (RTP) Stipend and RTP Fee-Offset Scholarship.
The GMRT is run by the National Centre for Radio Astrophysics of the Tata Institute of Fundamental Research, India. 

\end{acknowledgments}




\bibliography{sample631}{}

\begin{thebibliography}{}
\expandafter\ifx\csname natexlab\endcsname\relax\def\natexlab#1{#1}\fi
\providecommand{\url}[1]{\href{#1}{#1}}
\providecommand{\dodoi}[1]{doi:~\href{http://doi.org/#1}{\nolinkurl{#1}}}
\providecommand{\doeprint}[1]{\href{http://ascl.net/#1}{\nolinkurl{http://ascl.net/#1}}}
\providecommand{\doarXiv}[1]{\href{https://arxiv.org/abs/#1}{\nolinkurl{https://arxiv.org/abs/#1}}}

\bibitem[{Anumarlapudi {et~al.}(2023)Anumarlapudi, Swiggum, Kaplan, \&
  Fichtenbauer}]{Anumarlapudi2023}
Anumarlapudi, A., Swiggum, J.~K., Kaplan, D.~L., \& Fichtenbauer, T. D.~J.
  2023, \apj.
\newblock \doarXiv{2301.13258}

\bibitem[{{Backer}(1970)}]{Backer1970}
{Backer}, D.~C. 1970, \nat, 228, 42, \dodoi{10.1038/228042a0}

\bibitem[{{Basu} \& {Mitra}(2018)}]{basuMitra2018}
{Basu}, R., \& {Mitra}, D. 2018, \mnras, 476, 1345,
  \dodoi{10.1093/mnras/sty297}

\bibitem[{Basu {et~al.}(2017)Basu, Mitra, \& Melikidze}]{Basu2017}
Basu, R., Mitra, D., \& Melikidze, G.~I. 2017, The Astrophysical Journal, 846,
  109, \dodoi{10.3847/1538-4357/aa862d}

\bibitem[{Basu {et~al.}(2020)Basu, Mitra, \& Melikidze}]{Basu2020}
---. 2020, The Astrophysical Journal, 889, 133,
  \dodoi{10.3847/1538-4357/ab63c9}

\bibitem[{{Basu} {et~al.}(2016){Basu}, {Mitra}, {Melikidze}, {Maciesiak},
  {Skrzypczak}, \& {Szary}}]{Basu2016}
{Basu}, R., {Mitra}, D., {Melikidze}, G.~I., {et~al.} 2016, \apj, 833, 29,
  \dodoi{10.3847/1538-4357/833/1/29}

\bibitem[{{Basu} {et~al.}(2019{\natexlab{a}}){Basu}, {Mitra}, {Melikidze}, \&
  {Skrzypczak}}]{2019MNRAS.482.3757B}
{Basu}, R., {Mitra}, D., {Melikidze}, G.~I., \& {Skrzypczak}, A.
  2019{\natexlab{a}}, \mnras, 482, 3757, \dodoi{10.1093/mnras/sty2846}

\bibitem[{{Basu} {et~al.}(2019{\natexlab{b}}){Basu}, {Paul}, \&
  {Mitra}}]{Basu2019}
{Basu}, R., {Paul}, A., \& {Mitra}, D. 2019{\natexlab{b}}, \mnras, 486, 5216,
  \dodoi{10.1093/mnras/stz1225}

\bibitem[{{Bell} {et~al.}(2016){Bell}, {Murphy}, {Johnston}, {Kaplan}, {Croft},
  {Hancock}, {Callingham}, {Zic}, {Dobie}, {Swiggum}, {Rowlinson},
  {Hurley-Walker}, {Offringa}, {Bernardi}, {Bowman}, {Briggs}, {Cappallo},
  {Deshpande}, {Gaensler}, {Greenhill}, {Hazelton}, {Johnston-Hollitt},
  {Lonsdale}, {McWhirter}, {Mitchell}, {Morales}, {Morgan}, {Oberoi}, {Ord},
  {Prabu}, {Shankar}, {Srivani}, {Subrahmanyan}, {Tingay}, {Wayth}, {Webster},
  {Williams}, \& {Williams}}]{2016MNRAS.461..908B}
{Bell}, M.~E., {Murphy}, T., {Johnston}, S., {et~al.} 2016, \mnras, 461, 908,
  \dodoi{10.1093/mnras/stw1293}

\bibitem[{{Bhat} {et~al.}(2018){Bhat}, {Tremblay}, {Kirsten}, {Meyers},
  {Sokolowski}, {van Straten}, {McSweeney}, {Ord}, {Shannon}, {Beardsley},
  {Crosse}, {Emrich}, {Franzen}, {Horsley}, {Johnston-Hollitt}, {Kaplan},
  {Kenney}, {Morales}, {Pallot}, {Steele}, {Tingay}, {Trott}, {Walker},
  {Wayth}, {Williams}, \& {Wu}}]{2018ApJS..238....1B}
{Bhat}, N.~D.~R., {Tremblay}, S.~E., {Kirsten}, F., {et~al.} 2018, \apjs, 238,
  1, \dodoi{10.3847/1538-4365/aad37c}

\bibitem[{{Bhat} {et~al.}(2023{\natexlab{a}}){Bhat}, {Swainston}, {McSweeney},
  {Xue}, {Meyers}, {Kudale}, {Dai}, {Tremblay}, {van Straten}, {Shannon},
  {Smith}, {Sokolowski}, {Ord}, {Sleap}, {Williams}, {Hancock}, {Lange},
  {Tocknell}, {Johnston-Hollitt}, {Kaplan}, {Tingay}, \& {Walker}}]{smartone}
{Bhat}, N.~D.~R., {Swainston}, N.~A., {McSweeney}, S.~J., {et~al.}
  2023{\natexlab{a}}, \pasa, 40, e021, \dodoi{10.1017/pasa.2023.17}

\bibitem[{{Bhat} {et~al.}(2023{\natexlab{b}}){Bhat}, {Swainston}, {McSweeney},
  {Xue}, {Meyers}, {Kudale}, {Dai}, {Tremblay}, {van Straten}, {Shannon},
  {Smith}, {Sokolowski}, {Ord}, {Sleap}, {Williams}, {Hancock}, {Lange},
  {Tocknell}, {Johnston-Hollitt}, {Kaplan}, {Tingay}, \& {Walker}}]{smarttwo}
---. 2023{\natexlab{b}}, \pasa, 40, e020, \dodoi{10.1017/pasa.2023.18}

\bibitem[{{Biggs}(1992)}]{Biggs1992}
{Biggs}, J.~D. 1992, \apj, 394, 574, \dodoi{10.1086/171608}

\bibitem[{{Brentjens} \& {de Bruyn}(2005)}]{Brentjens2005}
{Brentjens}, M.~A., \& {de Bruyn}, A.~G. 2005, \aap, 441, 1217,
  \dodoi{10.1051/0004-6361:20052990}

\bibitem[{{Burn}(1966)}]{Burn1966}
{Burn}, B.~J. 1966, \mnras, 133, 67, \dodoi{10.1093/mnras/133.1.67}

\bibitem[{Caputo {et~al.}(2023)Caputo, Witte, Philippov, \&
  Jacobson}]{Caputo2023}
Caputo, A., Witte, S.~J., Philippov, A.~A., \& Jacobson, T. 2023, 1.
\newblock \doarXiv{arXiv:2311.14795v1}

\bibitem[{{Chen} \& {Ruderman}(1993)}]{1993ApJ...402..264C}
{Chen}, K., \& {Ruderman}, M. 1993, \apj, 402, 264, \dodoi{10.1086/172129}

\bibitem[{{CHIME/Pulsar Collaboration} {et~al.}(2021){CHIME/Pulsar
  Collaboration}, {Amiri}, {Bandura}, {Boyle}, {Brar}, {Cliche}, {Crowter},
  {Cubranic}, {Demorest}, {Denman}, {Dobbs}, {Dong}, {Fandino}, {Fonseca},
  {Good}, {Halpern}, {Hill}, {H{\"o}fer}, {Kaspi}, {Landecker}, {Leung}, {Lin},
  {Luo}, {Masui}, {McKee}, {Mena-Parra}, {Meyers}, {Michilli}, {Naidu},
  {Newburgh}, {Ng}, {Patel}, {Pinsonneault-Marotte}, {Ransom}, {Renard},
  {Scholz}, {Shaw}, {Sikora}, {Stairs}, {Tan}, {Tendulkar}, {Tretyakov},
  {Vanderlinde}, {Wang}, \& {Wang}}]{2021ApJS..255....5C}
{CHIME/Pulsar Collaboration}, {Amiri}, M., {Bandura}, K.~M., {et~al.} 2021,
  \apjs, 255, 5, \dodoi{10.3847/1538-4365/abfdcb}

\bibitem[{{Cordes} \& {Lazio}(2002)}]{ne2001}
{Cordes}, J.~M., \& {Lazio}, T.~J.~W. 2002, arXiv e-prints, astro,
  \dodoi{10.48550/arXiv.astro-ph/0207156}

\bibitem[{{Edwards} \& {Stappers}(2002)}]{2002A&A...393..733E}
{Edwards}, R.~T., \& {Stappers}, B.~W. 2002, \aap, 393, 733,
  \dodoi{10.1051/0004-6361:20021067}

\bibitem[{{Gajjar}(2017)}]{Gajjar2017}
{Gajjar}, V. 2017, arXiv e-prints, arXiv:1706.05407,
  \dodoi{10.48550/arXiv.1706.05407}

\bibitem[{{Han} {et~al.}(2018){Han}, {Manchester}, {van Straten}, \&
  {Demorest}}]{Han2018}
{Han}, J.~L., {Manchester}, R.~N., {van Straten}, W., \& {Demorest}, P. 2018,
  \apjs, 234, 11, \dodoi{10.3847/1538-4365/aa9c45}

\bibitem[{{Heald} {et~al.}(2009){Heald}, {Braun}, \& {Edmonds}}]{Heald2009}
{Heald}, G., {Braun}, R., \& {Edmonds}, R. 2009, \aap, 503, 409,
  \dodoi{10.1051/0004-6361/200912240}

\bibitem[{{Herfindal} \& {Rankin}(2007)}]{Herfindal2007}
{Herfindal}, J.~L., \& {Rankin}, J.~M. 2007, \mnras, 380, 430,
  \dodoi{10.1111/j.1365-2966.2007.12089.x}

\bibitem[{{Herfindal} \& {Rankin}(2009)}]{Herfindal2009}
---. 2009, \mnras, 393, 1391, \dodoi{10.1111/j.1365-2966.2008.14119.x}

\bibitem[{{Hobbs} {et~al.}(2004){Hobbs}, {Lyne}, {Kramer}, {Martin}, \&
  {Jordan}}]{Hobbs2004}
{Hobbs}, G., {Lyne}, A.~G., {Kramer}, M., {Martin}, C.~E., \& {Jordan}, C.
  2004, \mnras, 353, 1311, \dodoi{10.1111/j.1365-2966.2004.08157.x}

\bibitem[{{Janagal} {et~al.}(2022){Janagal}, {Chakraborty}, {Bhat},
  {Bhattacharyya}, \& {McSweeney}}]{Janagal2022}
{Janagal}, P., {Chakraborty}, M., {Bhat}, N.~D.~R., {Bhattacharyya}, B., \&
  {McSweeney}, S.~J. 2022, \mnras, 509, 4573, \dodoi{10.1093/mnras/stab3305}

\bibitem[{Kaplan {et~al.}(2018)Kaplan, Swiggum, Fichtenbauer, \&
  Vallisneri}]{Kaplan2018}
Kaplan, D.~L., Swiggum, J.~K., Fichtenbauer, T. D.~J., \& Vallisneri, M. 2018,
  The Astrophysical Journal, 855, 14, \dodoi{10.3847/1538-4357/aaab62}

\bibitem[{Konar \& Deka(2019)}]{Konar2019}
Konar, S., \& Deka, U. 2019, Journal of Astrophysics and Astronomy, 40,
  \dodoi{10.1007/s12036-019-9608-z}

\bibitem[{Kondratiev {et~al.}(2009)Kondratiev, McLaughlin, Lorimer, Burgay,
  Possenti, Turolla, Popov, \& Zane}]{Kondratiev2009}
Kondratiev, V.~I., McLaughlin, M.~A., Lorimer, D.~R., {et~al.} 2009,
  Astrophysical Journal, 702, 692, \dodoi{10.1088/0004-637X/702/1/692}

\bibitem[{{Manchester} {et~al.}(2005){Manchester}, {Hobbs}, {Teoh}, \&
  {Hobbs}}]{ATNF}
{Manchester}, R.~N., {Hobbs}, G.~B., {Teoh}, A., \& {Hobbs}, M. 2005, VizieR
  Online Data Catalog, VII/245

\bibitem[{{McSweeney} {et~al.}(2017){McSweeney}, {Bhat}, {Tremblay},
  {Deshpande}, \& {Ord}}]{McSweeney2017}
{McSweeney}, S.~J., {Bhat}, N.~D.~R., {Tremblay}, S.~E., {Deshpande}, A.~A., \&
  {Ord}, S.~M. 2017, \apj, 836, 224, \dodoi{10.3847/1538-4357/aa5c35}

\bibitem[{McSweeney {et~al.}(2022)McSweeney, Bhat, Swainston, Smith, Kudale,
  Hancock, Straten, Dai, Shannon, Tingay, Johnston-hollitt, Kaplan, \&
  Walker}]{Mcsweeney2022}
McSweeney, S.~J., Bhat, N. D.~R., Swainston, N.~A., {et~al.} 2022, The
  Astrophysical Journal, 933, 210, \dodoi{10.3847/1538-4357/ac75bc}

\bibitem[{Morello {et~al.}(2020)Morello, Barr, Stappers, Keane, \&
  Lyne}]{Morello2020}
Morello, V., Barr, E.~D., Stappers, B.~W., Keane, E.~F., \& Lyne, A.~G. 2020,
  Monthly Notices of the Royal Astronomical Society, 497, 4654,
  \dodoi{10.1093/mnras/staa2291}

\bibitem[{{Naidu} {et~al.}(2017){Naidu}, {Joshi}, {Manoharan}, \&
  {KrishnaKumar}}]{Naidu2017}
{Naidu}, A., {Joshi}, B.~C., {Manoharan}, P.~K., \& {KrishnaKumar}, M.~A. 2017,
  \aap, 604, A45, \dodoi{10.1051/0004-6361/201629937}

\bibitem[{{Ng} {et~al.}(2020){Ng}, {Wu}, {Ma}, {Ransom}, {Naidu}, {Fonseca},
  {Boyle}, {Brar}, {Cubranic}, {Demorest}, {Good}, {Kaspi}, {Masui},
  {Michilli}, {Patel}, {Renard}, {Scholz}, {Stairs}, {Tendulkar}, {Tretyakov},
  \& {Vanderlinde}}]{2020ApJ...903...81N}
{Ng}, C., {Wu}, B., {Ma}, M., {et~al.} 2020, \apj, 903, 81,
  \dodoi{10.3847/1538-4357/abb94f}

\bibitem[{{Radhakrishnan} \& {Cooke}(1969)}]{RVM1969}
{Radhakrishnan}, V., \& {Cooke}, D.~J. 1969, \aplett, 3, 225

\bibitem[{{Rahaman} {et~al.}(2021){Rahaman}, {Basu}, {Mitra}, \&
  {Melikidze}}]{Rahaman2021}
{Rahaman}, S. k.~M., {Basu}, R., {Mitra}, D., \& {Melikidze}, G.~I. 2021,
  \mnras, 500, 4139, \dodoi{10.1093/mnras/staa3518}

\bibitem[{{Rankin}(1983)}]{Rankin1983}
{Rankin}, J.~M. 1983, \apj, 274, 333, \dodoi{10.1086/161450}

\bibitem[{{Rankin}(1986)}]{Rankin1986}
---. 1986, \apj, 301, 901, \dodoi{10.1086/163955}

\bibitem[{{Ritchings}(1976)}]{Ritchings1976}
{Ritchings}, R.~T. 1976, \mnras, 176, 249, \dodoi{10.1093/mnras/176.2.249}

\bibitem[{Ruderman \& Sutherland(1975)}]{Ruderman1975}
Ruderman, M.~A., \& Sutherland, P.~G. 1975, Astrophysical Journal, 4, 10

\bibitem[{{Sanidas} {et~al.}(2019){Sanidas}, {Cooper}, {Bassa}, {Hessels},
  {Kondratiev}, {Michilli}, {Stappers}, {Tan}, {van Leeuwen}, {Cerrigone},
  {Fallows}, {Iacobelli}, {Orr{\'u}}, {Pizzo}, {Shulevski}, {Toribio}, {ter
  Veen}, {Zucca}, {Bondonneau}, {Grie{\ss}meier}, {Karastergiou}, {Kramer}, \&
  {Sobey}}]{2019A&A...626A.104S}
{Sanidas}, S., {Cooper}, S., {Bassa}, C.~G., {et~al.} 2019, \aap, 626, A104,
  \dodoi{10.1051/0004-6361/201935609}

\bibitem[{Sheikh \& Macdonald(2021)}]{Sheikh2021}
Sheikh, S.~Z., \& Macdonald, M.~G. 2021, Monthly Notices of the Royal
  Astronomical Society, 502, 4669, \dodoi{10.1093/mnras/stab282}

\bibitem[{{Sun} {et~al.}(2021){Sun}, {Yan}, \& {Wang}}]{2021MNRAS.501.3900S}
{Sun}, S.~N., {Yan}, W.~M., \& {Wang}, N. 2021, \mnras, 501, 3900,
  \dodoi{10.1093/mnras/staa3825}

\bibitem[{{Swainston} {et~al.}(2022{\natexlab{a}}){Swainston}, {Bhat},
  {Morrison}, {McSweeney}, {Ord}, {Tremblay}, \&
  {Sokolowski}}]{2022PASA...39...20S}
{Swainston}, N.~A., {Bhat}, N.~D.~R., {Morrison}, I.~S., {et~al.}
  2022{\natexlab{a}}, \pasa, 39, e020, \dodoi{10.1017/pasa.2022.14}

\bibitem[{{Swainston} {et~al.}(2022{\natexlab{b}}){Swainston}, {Bhat},
  {Morrison}, {McSweeney}, {Ord}, {Tremblay}, \& {Sokolowski}}]{swainston2022}
---. 2022{\natexlab{b}}, \pasa, 39, e020, \dodoi{10.1017/pasa.2022.14}

\bibitem[{{Swainston} {et~al.}(2021){Swainston}, {Bhat}, {Sokolowski},
  {McSweeney}, {Kudale}, {Dai}, {Smith}, {Morrison}, {Shannon}, {van Straten},
  {Xue}, {Ord}, {Tremblay}, {Meyers}, {Williams}, {Sleap}, {Johnston-Hollitt},
  {Kaplan}, {Tingay}, \& {Wayth}}]{psrone}
{Swainston}, N.~A., {Bhat}, N.~D.~R., {Sokolowski}, M., {et~al.} 2021, \apjl,
  911, L26, \dodoi{10.3847/2041-8213/abec7b}

\bibitem[{{Tingay} {et~al.}(2013){Tingay}, {Goeke}, {Bowman}, {Emrich}, {Ord},
  {Mitchell}, {Morales}, {Booler}, {Crosse}, {Wayth}, {Lonsdale}, {Tremblay},
  {Pallot}, {Colegate}, {Wicenec}, {Kudryavtseva}, {Arcus}, {Barnes},
  {Bernardi}, {Briggs}, {Burns}, {Bunton}, {Cappallo}, {Corey}, {Deshpande},
  {Desouza}, {Gaensler}, {Greenhill}, {Hall}, {Hazelton}, {Herne}, {Hewitt},
  {Johnston-Hollitt}, {Kaplan}, {Kasper}, {Kincaid}, {Koenig}, {Kratzenberg},
  {Lynch}, {Mckinley}, {Mcwhirter}, {Morgan}, {Oberoi}, {Pathikulangara},
  {Prabu}, {Remillard}, {Rogers}, {Roshi}, {Salah}, {Sault}, {Udaya-Shankar},
  {Schlagenhaufer}, {Srivani}, {Stevens}, {Subrahmanyan}, {Waterson},
  {Webster}, {Whitney}, {Williams}, {Williams}, \& {Wyithe}}]{Tingay2013}
{Tingay}, S.~J., {Goeke}, R., {Bowman}, J.~D., {et~al.} 2013, \pasa, 30, e007,
  \dodoi{10.1017/pasa.2012.007}

\bibitem[{Wang {et~al.}(2007)Wang, Manchester, \& Johnston}]{Wang2007}
Wang, N., Manchester, R.~N., \& Johnston, S. 2007, Monthly Notices of the Royal
  Astronomical Society, 377, 1383, \dodoi{10.1111/j.1365-2966.2007.11703.x}

\bibitem[{{Wayth} {et~al.}(2018){Wayth}, {Tingay}, {Trott}, {Emrich},
  {Johnston-Hollitt}, {McKinley}, {Gaensler}, {Beardsley}, {Booler}, {Crosse},
  {Franzen}, {Horsley}, {Kaplan}, {Kenney}, {Morales}, {Pallot}, {Sleap},
  {Steele}, {Walker}, {Williams}, {Wu}, {Cairns}, {Filipovic}, {Johnston},
  {Murphy}, {Quinn}, {Staveley-Smith}, {Webster}, \&
  {Wyithe}}]{2018PASA...35...33W}
{Wayth}, R.~B., {Tingay}, S.~J., {Trott}, C.~M., {et~al.} 2018, \pasa, 35,
  e033, \dodoi{10.1017/pasa.2018.37}

\bibitem[{Weltevrede {et~al.}(2006)Weltevrede, Edwards, \&
  Stappers}]{Weltevrede2006}
Weltevrede, P., Edwards, R.~T., \& Stappers, B.~W. 2006, Astronomy and
  Astrophysics, 445, 243, \dodoi{10.1051/0004-6361:20053088}

\bibitem[{{Wen} {et~al.}(2016){Wen}, {Wang}, {Yuan}, {Yan}, {Manchester},
  {Yuen}, \& {Gajjar}}]{Wen2016}
{Wen}, Z.~G., {Wang}, N., {Yuan}, J.~P., {et~al.} 2016, \aap, 592, A127,
  \dodoi{10.1051/0004-6361/201628214}

\bibitem[{{Wen} {et~al.}(2020){Wen}, {Yan}, {Yuan}, {Wang}, {Chen}, {Mijit},
  {Yuen}, {Wang}, {Tu}, \& {Dang}}]{2020Wen_modechange}
{Wen}, Z.~G., {Yan}, W.~M., {Yuan}, J.~P., {et~al.} 2020, \apj, 904, 72,
  \dodoi{10.3847/1538-4357/abbfa3}

\bibitem[{{Wen} {et~al.}(2022){Wen}, {Yuan}, {Wang}, {Li}, {Chen}, {Wang},
  {Wu}, {Yan}, {Yuen}, {Wang}, {Tedila}, {Wang}, {Zhu}, {Niu}, {Miao}, {Xue},
  {Duan}, {Xiang}, \& {He}}]{Wen2022}
{Wen}, Z.~G., {Yuan}, J.~P., {Wang}, N., {et~al.} 2022, \apj, 929, 71,
  \dodoi{10.3847/1538-4357/ac5d5d}

\bibitem[{{Xue} {et~al.}(2019){Xue}, {Ord}, {Tremblay}, {Bhat}, {Sobey},
  {Meyers}, {McSweeney}, \& {Swainston}}]{2019PASA...36...25X}
{Xue}, M., {Ord}, S.~M., {Tremblay}, S.~E., {et~al.} 2019, \pasa, 36, e025,
  \dodoi{10.1017/pasa.2019.19}

\bibitem[{{Yao} {et~al.}(2017){Yao}, {Manchester}, \& {Wang}}]{ymw16}
{Yao}, J.~M., {Manchester}, R.~N., \& {Wang}, N. 2017, \apj, 835, 29,
  \dodoi{10.3847/1538-4357/835/1/29}

\bibitem[{Zhang {et~al.}(2000)Zhang, Harding, \& Muslimov}]{Zhang2000}
Zhang, B., Harding, A.~K., \& Muslimov, A.~G. 2000, The Astrophysical Journal,
  531, L135, \dodoi{10.1086/312542}

\end{thebibliography}
\bibliographystyle{aasjournal}




\end{document}